\documentclass[a4paper,11pt]{article}
\pdfoutput=1 
\usepackage{jcappub}
\usepackage{amsmath}
\usepackage{graphicx}
\usepackage{latexsym}
\usepackage{xspace}
\usepackage{color}
\usepackage{hyperref} 
\usepackage{bm}
\usepackage{relsize}
\usepackage{tabularx}
\usepackage{multirow}
\usepackage{amssymb}
\usepackage[table]{xcolor}
\usepackage{braket}
\usepackage{comment}
\usepackage[utf8]{inputenc}
\usepackage{slashed}

\makeatletter
\gdef\@fpheader{}
\g@addto@macro\bfseries{\boldmath}
\makeatother

\newcommand{\ds}{\displaystyle}

\newcommand{\ie}{\textsl{i.e.~}}

\newcommand{\eg}{\textsl{e.g.~}}

\newcommand{\order}[1]{\mathcal{O}\!\left(#1\right)}

\newcommand{\dd}{\mathrm{d}}
\newcommand{\ee}{e}

\newcommand{\sss}[1]{{\scriptscriptstyle{#1}}}
\newcommand{\boldmathsymbol}[1]{{\ensuremath{\boldsymbol{#1}}}}

\newcommand{\uPl}{\mathrm{Pl}}
\newcommand{\uin}{\mathrm{in}}

\newcommand{\usssPl}{\sss{\uPl}}

\newcommand{\calP}{\mathcal{P}}

\newcommand{\Mp}{M_\usssPl}

\newcommand{\beq}{\begin{equation}}
\newcommand{\eeq}{\end{equation}}
\newcommand{\bea}{\begin{eqnarray}}
\newcommand{\eea}{\end{eqnarray}}

\newlength{\wsingfig}
\newlength{\wdblefig}
\newlength{\wquadfig}
\newlength{\wtriplefig}
\newcommand{\Eq}[1]{Eq.~(\ref{#1})}
\newcommand{\Eqs}[1]{Eqs.~(\ref{#1})}
\newcommand{\Fig}[1]{Fig.~{\ref{#1}}}

\newcommand{\Refc}[1]{Ref.~{\cite{#1}}}
\newcommand{\Refs}[1]{Refs.~{\cite{#1}}}
\newcommand{\Sec}[1]{Sec.~\ref{#1}}

\newcommand{\App}[1]{Appendix~\ref{#1}}

\subheader{}

\title{Unavoidable shear from quantum fluctuations in contracting cosmologies}

\author[a]{Julien Grain,}
\affiliation[a]{Universit\'e Paris-Saclay, CNRS, Institut d'Astrophysique Spatiale, 91405, Orsay, France}
\emailAdd{julien.grain@universite-paris-saclay.fr}

\author[b]{Vincent Vennin}
\affiliation[b]{Laboratoire Astroparticule et Cosmologie, CNRS, Universit\'e de Paris, 75013 Paris, France}
\emailAdd{vincent.vennin@apc.univ-paris7.fr}

\date{today}

\begin{document}
\sloppy

\abstract{Contracting cosmologies are known to be flawed with a shear instability, where the contribution from the anisotropic stress to the overall energy density grows as $a^{-6}$, with $a$ the scale factor. Classically, whether or not this contribution becomes important before the bounce depends on its initial value, which can always be sufficiently fine tuned to make it irrelevant. However, vacuum quantum fluctuations inevitably provide a non-vanishing source of anisotropic stress. In this work, we compute the minimum amount of shear that is obtained if one assumes that it vanishes initially, but lets quantum fluctuations build it up. In practice, we consider a massless test scalar field, and describe its quantum fluctuations by means of the stochastic ``inflation'' (though here applied to a contracting phase) formalism. We find that, if the equation-of-state parameter of the contraction satisfies $w>-1/9$, regardless of when the contracting phase is initiated, the time at which the shear becomes sizeable is always when the Hubble scale approaches the Planck mass (which is also where the bounce is expected to take place). However, if $w<-1/9$, the shear backreaction becomes important much earlier, at a point that depends on the overall amount of contraction.}



\maketitle


%
\section{Introduction}
\label{sec:intro}
Although inflation is the leading paradigm to describe the early universe, it is sometimes assumed to be preceded by a phase of slow contraction, followed by a bounce. This is indeed naturally expected in most theories of quantum gravity, it avoids the initial singularity that is otherwise still present in inflation when considered alone, and can also solve the trans-Planckian problem~(see \eg \Refs{Battefeld:2014uga,Brandenberger:2016vhg,Agullo:2016tjh} and references therein). In a homogeneous cosmology, the energy density associated with the anisotropic stress however grows as $a^{-6}$, where $a$ is the scale factor. Since the background energy density grows as $a^{-3(1+w)}$, where $w$ is the equation-of-state parameter of the background fluid, any initial anisotropy grows unstable as the universe contracts, hence driving a new kind of singularity~\cite{Belinsky:1970ew}, unless $w>1$.

Several solutions to this problem have been proposed. One can for instance assume that the universe contracts with a stiff equation of state, $w>1$, preventing the energy density associated with the anisotropic stress to overcome the isotropic contribution. This corresponds for instance to the ekpyrotic model~\cite{Khoury:2001wf}. Another solution more rooted in quantum gravity consists in regularising the space-time curvature. A concrete example of this is Loop Quantum Cosmology of anisotropic models of the universe \cite{Ashtekar_2009}. Here the quantisation of the gravitational degrees of freedom makes all energy densities (including the one associated with the anisotropic stress) effectively bounded from above, then guaranteeing a regular bounce for any equation of state.

Obviously, a third solution is to fine tune the initial anisotropy to be sufficiently small such that it remains negligible until the bounce occurs. The amount of required fine tuning may be large for a long phase of contraction, but, at the classical level, it is always possible to set the initial anisotropy to arbitrarily small values. However, in the presence of vacuum quantum fluctuations, this is not true anymore and there is a minimum, inevitable amount of anisotropic stress sourced by quantum fluctuations. 

The goal of this work is to study this unavoidable shear production sourced by quantum fluctuations (note that vector perturbations in bouncing cosmologies were also studied at the classical level in \Refc{Pinto-Neto:2020xmb}).
 It is organised as follows. In \Sec{sec:FormulationProblem}, we formulate the problem, and show why quantum fluctuations of a test scalar field provide a non-vanishing source of anisotropic stress (more precisely, we explain why the one-point function of the anisotropic stress produced by scalar-field fluctuations vanish, but not its two-point function). In \Sec{sec:Stochastic:Formalism}, we derive a stochastic formalism, analogous to the stochastic-inflation formalism, to describe these quantum fluctuations in an inflating or a contracting cosmology, and to compute the expectation value of the shear they generate. In \Sec{sec:Contraction}, we apply these tools to the case of a contracting phase with a constant equation-of-state parameter $w$. We find that, when $w>-1/9$, the shear produced out of massless scalar field fluctuations is always negligible until the energy density of the universe reaches the Planck scale, regardless of initial conditions; while for $w<-1/9$, it becomes sizeable much before then, at a point that depends on the time at which the contraction is initiated. In \Sec{sec:Discussion}, we compare the shear backreaction with the direct contribution of the test scalar field fluctuations to the energy density of the universe, and conclude that the later is always more important than the former. We also discuss the case of inflating backgrounds, where it is found that the shear backreaction is always negligible. We present our conclusions in \Sec{sec:Conclusion}.
\section{Formulating the problem}
\label{sec:FormulationProblem}
We consider the case where deviations from the isotropic Friedmann-Lema\^itre-Robertson-Walker (FLRW) space-time can be parametrised by a Bianchi I metric~\cite{Ellis:1968vb, Pereira:2007yy},
\bea
\label{eq:metric:Bianchi}
	\dd s^2=-\dd t^2+a^2(t)\gamma_{ij}\dd x^i\dd x^j,
\eea
with $a(t)$ the isotropised scale factor and $\gamma_{ij}(t)=\exp[2\beta_i(t)]\delta_{ij}$. The $\beta_i$ functions are constrained to satisfy $\sum_i\beta_i=0$, such that $\det \gamma_{ij}=1$. The shear tensor, $\sigma_{ij}$, is defined as 
\bea
\sigma_{ij}=\frac{\dot\gamma_{ij}}{2}\, ,
\eea
where a dot denotes a derivative with respect to cosmic time $t$. 

We further assume that the universe is filled with a homogeneous but anisotropic matter fluid, described by its energy density $\rho$, its pressure $P$ and its anisotropic stress $\Pi_{\mu\nu}$, such that its stress-energy tensor reads
\bea
\label{eq:Tmunu:fluid}
T_{\mu\nu} = \left(\rho+P\right) u_{\mu}u_{\nu} + P g_{\mu\nu} +  \Pi_{\mu\nu}\, .
\eea 
In this expression, $u^\mu$ is the four velocity of a comoving observer, normalised such that $u_\mu u^\mu=-1$. From the metric~\eqref{eq:metric:Bianchi}, a comoving observer has $u^i=0$, and the normalisation condition imposes that $u^\mu=-\delta^\mu_0$. The anisotropic stress is a symmetric tensor (since the stress-energy tensor is symmetric) that parametrises deviations from a perfect fluid. Given that its ``$00$'' component can always be reabsorbed in the first term of \Eq{eq:Tmunu:fluid}, hence in the definition of $\rho$, it is set to zero without loss of generality. The ``$0i$'' component of the Einstein equation then imposes that $\Pi_{0i}=0$, which implies that the anisotropic stress is orthogonal to the comoving velocity, $\Pi_{\mu\nu} u^\nu=0$. If one decomposes $\Pi_{ij}$ as $\Pi_{ij}=\Pi_{ij}-  g^{mn}\Pi_{mn} g_{ij}/3 + g^{mn}\Pi_{mn}g_{ij} /3$, the component $  g^{mn}\Pi_{mn}g_{ij} /3$ can always be reabsorbed in the second term of \Eq{eq:Tmunu:fluid}, hence in the definition of $P$, and the remaining part $\Pi_{ij}-  g^{mn}\Pi_{mn} g_{ij}/3$ is traceless, so we assume that $\Pi_{ij}$ is traceless without loss of generality, $\Pi_{ij} g^{ij}=0$. 

Under these conditions, the Einstein equations can be written as
\bea
\label{eq:Friedmann}
H^2& = & \frac{\rho}{3\Mp^2}+\frac{1}{6} \sigma_{ij} \sigma^{ij}\, ,\\
\label{eq:Raychaudhuri}
\dot{H} & = & -\frac{\rho+P}{2\Mp^2}-\frac{1}{2}  \sigma_{ij} \sigma^{ij}\, ,\\
\label{eq:sigmadot:Pi}
\left(\sigma^i_j\right){\dot{}}+3 H \sigma^i_j &=& \frac{\Pi^i_j}{\Mp^2}\, ,
\eea
which correspond respectively to the ``$00$'' component, the trace and the traceless part of the ``$ij$'' component, and where $H\equiv\dot{a}/a$ is the Hubble parameter. By combining these three equations, one obtains a conservation relation,
\bea
\label{eq:conservation:equation}
\dot{\rho}+3H\left(\rho+P\right)=-\sigma_{ij}\Pi^{ij}\, .
\eea
At the background level, one can see that the only effect of the shear is to add a contribution $\rho_\sigma = \Mp^2  \sigma_{ij} \sigma^{ij}/2$ to the energy density, and $p_\sigma = \rho_\sigma$ to the pressure, in the Friedmann equation~\eqref{eq:Friedmann} and in the Raychaudhuri equation~\eqref{eq:Raychaudhuri}. This additional contribution is sourced by the anisotropic stress through \Eq{eq:sigmadot:Pi}, which can be formally solved as
\bea
\label{eq:sigma:sol}
\sigma^i_j (t) = \sigma^i_j \left(t_\uin\right)\left[\frac{a\left(t_\uin\right)}{a\left(t\right)}\right]^3 + \frac{1}{\Mp^2}\int_{t_\uin}^t \left[\frac{a\left(t^\prime\right)}{a\left(t\right)}\right]^3 \Pi^{i}_{j}\left(t^\prime\right)\dd t^\prime\, .
\eea
From this expression, one can see that, in the absence of anisotropic stress, any initial amount of shear tensor grows as $\sigma^i_{j} \propto a^{-3}$, which gives rise to $\rho_\sigma\propto a^{-6}$. Denoting by $w$ the equation-of-state parameter of the background fluid, such that $P=w\rho$, \Eq{eq:conservation:equation} gives rise to $\rho\propto a^{-3(1+w)}$ in the absence of anisotropic stress, so the shear contribution eventually dominates unless $w>1$, as mentioned above.

The time at which $\rho_\sigma$ takes over $\rho$ obviously depends on its initial value, hence on $\sigma^i_j(t_\uin)$. Since the contracting phase ends at the time when the universe experiences a bounce, and given that the bounce is followed by an expanding phase, during which $\rho_\sigma$ decreases faster than $\rho$ (unless $w>1$), it is always possible to make the shear irrelevant by requiring its initial value to be sufficiently small. The shear problem thus becomes an initial fine-tuning problem. There is, however, no natural choice for neither the values of the shear tensor nor the probabilities associated to them, and this could have important consequences for the state of the universe after the bounce (see \Refs{Linsefors:2013bua, Linsefors:2014tna} for a concrete example in Loop Quantum Cosmology). Alternatively one could imagine the existence of mechanisms in the early stage of, or prior to, the contracting phase, that would set the shear to the required small value, as could be the case in cyclic models of the universe. 

These considerations are however valid only under the assumption that the anisotropic stress vanishes. While this is true in principle for a perfect fluid, at the high energy scales that operate in the early universe, field theory is the relevant framework to describe matter, and as we shall see, the unavoidable vacuum fluctuations present in quantum fields are enough to generate an anisotropic stress. For explicitness, let us consider the case of a test scalar field $\phi$, minimally coupled to gravity, described by the action
\bea
\label{eq:action:phi}
S^{(\phi)} = -\int\dd^4 \bm{x} \sqrt{-g} \left[\frac{1}{2}g^{\mu\nu}\partial_\mu\phi\partial_\nu\phi + V\left(\phi\right)\right],
\eea
where $V(\phi)$ is the potential associated with $\phi$. The stress-energy tensor can be obtained from $T_{\mu\nu}^{(\phi)} = -(2/\sqrt{-g})   \delta S^{(\phi)}/\delta g^{\mu\nu} $, which leads to
\bea
\label{eq:StressEnergyTensor:ScalarField}
T_{\mu\nu}^{(\phi)} = \partial_\mu\phi \partial_\nu\phi 
-  \left[ \frac{1}{2} g^{\kappa\lambda}\partial_\kappa\phi \partial_\lambda\phi + V\left(\phi\right) \right] g_{\mu\nu}\, .
\eea
As explained above, the anisotropic stress corresponds to the transverse and traceless part of the stress-energy tensor. Transverseness is obtained by projecting $T_{\mu\nu}^{(\phi)}$ in the hyperplane orthogonal to $u^\mu$, by using the projector $\perp_{\mu\nu}=g_{\mu\nu}+u_\mu u_\nu$.\footnote{From the normalisation condition of $u^\mu$, $u_\mu u^\mu=-1$, one can indeed check that (i) $\perp_{\mu\nu} u^\nu =0$, (ii) if $v^\mu$ is a vector orthogonal to $u^\mu$, \ie $u_\mu v^\mu=0$, then $\perp_{\mu\nu} v^\nu = v_\mu$, and (iii) $\perp_{\mu}^\rho \perp_{\rho\sigma} = \perp_{\mu\nu}$.} In the Bianchi I space time~\eqref{eq:metric:Bianchi}, its only non-zero components are $\perp_{ij}=g_{ij}$, which is no more than the induced metric. Tracelessness is obtained by removing the trace of the projected tensor, and one obtains $\Pi_{ij}^{(\phi)} = T^{(\phi)}_{ij }-\frac{1}{3}g_{ij}g^{mn} T_{mn}^{(\phi)}$, hence
\bea
\label{eq:AnisotropicStress:ScalarField}
\Pi_{ij}^{(\phi)} = \partial_i \phi \partial_j \phi -\frac{1}{3}\left(g^{mn}\partial_m\phi\partial_n\phi \right) g_{ij}\, .
\eea
Let us consider the case where the initial shear vanishes, so space time is initially isotropic. If the quantum state of the scalar field $\phi$ shares the same symmetries as the background, for instance if it is in its vacuum state $\vert 0 \rangle$, then there is no preferred direction and $\langle 0 \vert \Pi_{ij}^{(\phi)}  \vert  0 \rangle =0 $ (this will be shown explicitly below). This implies that, when taking the expectation value of \Eq{eq:sigma:sol}, the anisotropic stress disappears, so $\langle \sigma^i_{j} \rangle \propto a^{-3}$ and the above considerations still apply. However, the two-point function of the anisotropic stress does not necessarily vanish, and since the self contracted-shear appears in $\rho_\sigma$, this means that quantum fluctuations of a scalar field can source an effective shear. In other words, by self contracting \Eq{eq:sigma:sol}, in the absence of initial shear, one has
\bea
\label{eq:rhoSigma:AnisotropicStress}
\rho_\sigma = \frac{1}{2\Mp^2} 
\int_{t_\uin}^t \dd t_1\int_{t_\uin}^t \dd t_2 \left[\frac{a\left(t_1\right)a\left(t_2\right)}{a^2(t)}\right]^3 \Pi_{i}^{j} \left(t_1\right) \Pi^{i}_{j}\left(t_2\right)\, .
\eea
The goal of the following is thus to compute the two-point function of the anisotropic stress for a test scalar field, in order to evaluate \Eq{eq:rhoSigma:AnisotropicStress}. Let us note that, since $\phi$ must be homogeneous at the background level, \Eq{eq:AnisotropicStress:ScalarField} is of second order in the field perturbations, hence $\rho_\sigma$ is of fourth order in those perturbations. This small contribution is however enhanced by $a^{-6}$, so the (ir)relevance of the effect is a priori not obvious. Let us also note that other types of fields, such as vector fields or gauge fields, can give a non-vanishing mean anisotropic stress, hence source the shear at second rather than fourth order in perturbations. In this work, we choose to study the case of a scalar field, in order to extract the minimum amount of shear one cannot avoid, simply by postulating the existence of light scalar degrees of freedom during the contraction.

A last word is in order regarding the strategy followed in this work. Hereafter, the question we are interested in is whether or not a homogeneous and isotropic contracting cosmology is stable under quantum fluctuations of a test scalar field. In practice, the stability analysis is performed in a specific ``direction'', namely the would-be contribution to the anisotropic stress coming from a test scalar field, in a Bianchi I model. This is because, at the classical level, the shear instability signals the presence of a potential danger in this direction. However, this does not allow us to draw conclusions about stability in other ``directions'' (this would for instance require to go beyond the Bianchi I space-time). Moreover, our formalism is not meant to describe backreaction, but rather to provide an indicator of whether or not backreaction might become important. This is done by computing an effective energy density associated to the anisotropic stress, and by comparing this quantity to the overall energy density of the background, while still assuming that the space-time geometry is fixed, and close to isotropic. If one found situations where backreaction becomes important, then one would also have to include fluctuations in the metric sector (which are known to couple in anisotropic universes~\cite{Pereira:2007yy}) to describe the full dynamics of the problem. This however goes beyond the scope of this paper.

\section{Stochastic shear from a test scalar field}
\label{sec:Stochastic:Formalism}
In order to compute the statistics of the anisotropic stress generated by a scalar field in a contracting phase, we employ a stochastic formalism reminiscent of the ``stochastic inflation'' formalism~\cite{Starobinsky:1982ee, Starobinsky:1986fx, Starobinsky:1994bd}. While this makes practical calculations easier, a comparison with a full quantum field theoretic treatment is performed in \App{sec:app:QFT}, where we check that the ``stochastic contraction'' approach developed in this work is indeed valid (a result that is well-established in the context of inflation, see \eg \Refs{Finelli:2008zg, Garbrecht:2013coa, Vennin:2015hra}). Another reason for using a stochastic formalism is that, although only the case of a test field is considered hereafter, it may help to generalise our results to describe the backreaction of this field on the background dynamics, see the discussion in \Sec{sec:Conclusion}. 
\subsection{Langevin equations for a test scalar field}
\label{sec:Langevin:Free:Scalar}
During inflation, because of the presence of a phase-space attractor, the so-called slow-roll solution~\cite{Salopek:1990jq, Liddle:1994dx}, the stochastic formalism reduces to a single Langevin equation once this attractor is reached. In a contracting cosmology, there is no such attractor and the stochastic formalism has to be extended to the full phase space, as was done in \Refc{Grain:2017dqa}, which we follow below. Starting from \Eq{eq:action:phi}, and working with cosmic time,\footnote{In general, the choice of the time coordinate with which the Langevin equations are derived determines the gauge in which the calculation is performed~\cite{Pattison:2019hef} (namely the synchronous gauge for that time variable). When the field perturbations couple to metric perturbations, this sets the gauge in which the noise needs to be computed. However, in the present case of a test field, such a coupling is neglected and there is no subtlety associated with the choice of time coordinate.} the Hamilton equations read
\bea
\label{eq:eom:phi}
	\dot\phi&=&\frac{\pi}{a^3}\, , \\
	\dot\pi&=&a(t)e^{-2\beta_i(t)}\delta^{ij}\partial_i\partial_j\phi -a^3 m^2 \phi\, .
\label{eq:eom:pi}
\eea
Hereafter, for simplicity, we consider the case of a free field with potential $V(\phi)=m^2\phi^2/2$.
The coarse-graining procedure then consists in splitting the scalar field into a large-scale (classical) part and small-scale (quantum) fluctuations as $\phi=\bar\phi+\phi_Q$ and $\pi=\bar\pi+\pi_Q$, with
\bea
\label{eq:phiQ}
	\hat{\phi}_Q&=&\ds\int\frac{\dd^3\vec{k}}{(2\pi)^{3/2}}W\left(\vec{k},t\right)\left[\hat{a}_{\vec{k}}\phi_{\vec{k}}(t)e^{-i\vec{k}\cdot\vec{x}}+\hat{a}_{\vec{k}}^\dagger\phi_{\vec{k}}^*(t)e^{i\vec{k}\cdot\vec{x}}\right], \\
	\hat{\pi}_Q&=&\ds\int\frac{\dd^3\vec{k}}{(2\pi)^{3/2}}W\left(\vec{k},t\right)\left[\hat{a}_{\vec{k}}\pi_{\vec{k}}(t)e^{-i\vec{k}\cdot\vec{x}}+\hat{a}_{\vec{k}}^\dagger\pi_{\vec{k}}^*(t)e^{i\vec{k}\cdot\vec{x}}\right].
\label{eq:piQ}
\eea
Here, $W$ is a window function that selects out small-wavelength modes, \ie it is such that $W \simeq 1$ for $k\gg k_\epsilon(\vec{k}/k,t)$ and $W\simeq 0$ otherwise. In an FLRW background, the scale of coarse-graining is chosen as $k_\epsilon=\epsilon a \vert H\vert$ with $\epsilon\ll1$, for two reasons. First, this ensures that the coarse-grained part is indeed comprised of large-scale modes only, well above the Hubble radius, where gradients can be neglected (see below and the discussion in \App{sec:app:QFT}). Second, this guarantees that the inflow of modes into the large-scale sector is made of quantum fluctuations that are sufficiently squeezed and amplified to be considered as a classical stochastic noise. Here, since the background is anisotropic, the coarse-graining scale not only depends on time but also on the direction of $\vec{k}$, which is why the window function depends both on the norm and direction of $\vec{k}$. In each spatial direction, from \Eq{eq:metric:Bianchi} one can define a scale factor $a_i=a e^{\beta_i}$, hence a Hubble rate $H_i = H + \dot{\beta}_i$. If one defines the coarse-grained field as being made of wavelengths that are much larger than the Hubble radius in \emph{all} directions,
\bea
k_i<\epsilon a_i \left\vert H_i\right\vert  = \epsilon a e^{\beta_i} \left\vert H+\dot{\beta}_i\right\vert\ \forall i,
\eea
where $\epsilon\ll 1$, this gives a prescription for $ k_\epsilon(\vec{k}/k,t)$. 

In \Eqs{eq:phiQ} and~\eqref{eq:piQ}, $\hat{a}^\dagger_{\vec{k}}$ and $\hat{a}_{\vec{k}}$ are creation and annihilation operators, that satisfy the canonical commutation relations $[a_{\vec{k}},a^\dag_{\vec{q}}]=\delta^3(\vec{k}-\vec{q})$. The mode functions $\phi_{\vec{k}}$ and $\pi_{\vec{k}}$, which satisfy the equations of motion~\eqref{eq:eom:phi} and~\eqref{eq:eom:pi} when written in Fourier space, have to be normalised according to the Klein-Gordon product $i\int_{\Sigma_t}\dd^3x(\Phi_{\vec{k}}\Pi^*_{\vec{q}}-\Pi_{\vec{k}}\Phi^*_{\vec{q}})=\delta^3(\vec{k}-\vec{q})$, where $\Phi_{\vec{k}}=\phi_{\vec{k}}(t)e^{-i\vec{k}\cdot\vec{x}}$ and $\Pi_{\vec{k}}=\pi_{\vec{k}}(t)e^{-i\vec{k}\cdot\vec{x}}$, and where $\Sigma_t$ is a space-like hypersurface at fixed time $t$.

Inserting the field decomposition $\phi=\bar\phi+\phi_Q$ and $\pi=\bar\pi+\pi_Q$ into \Eqs{eq:eom:phi} and~\eqref{eq:eom:pi}, one obtains Langevin equations for the coarse-grained fields,
\bea
\label{eq:Langevin:phi}
\dot{\bar{\phi}}&=&\frac{\bar{\pi}}{a^3}+\xi_\phi\, ,\\
\dot{\bar{\pi}}&=&-a^3 m^2 \bar{\phi} + \xi_\pi\, ,
\label{eq:Langevin:pi}
\eea
where gradient terms are neglected since they are suppressed by $\epsilon^2$. The two noise terms $\xi_\phi$ and $\xi_\pi$ are random Gaussian noises, that share the same statistics as their quantum counterparts
\bea
\label{eq:xi:phi}
	\hat{\xi}_\phi&=&\ds -\int\frac{\dd^3\vec{k}}{(2\pi)^{3/2}}\frac{\partial W(\vec{k},t)}{\partial t}\left[\hat{a}_{\vec{k}}\phi_{\vec{k}}(t)e^{-i\vec{k}\cdot\vec{x}}+\hat{a}_{\vec{k}}^\dagger\phi_{\vec{k}}^*(t)e^{i\vec{k}\cdot\vec{x}}\right], \\
	\hat{\xi}_\pi&=&\ds - \int\frac{\dd^3\vec{k}}{(2\pi)^{3/2}}\frac{\partial W(\vec{k},t)}{\partial t}\left[\hat{a}_{\vec{k}}\pi_{\vec{k}}(t)e^{-i\vec{k}\cdot\vec{x}}+\hat{a}_{\vec{k}}^\dagger\pi_{\vec{k}}^*(t)e^{i\vec{k}\cdot\vec{x}}\right].
\label{eq:xi:pi}
\eea
At leading order in perturbation theory, this statistics is Gaussian, hence fully characterised by its two-point correlation functions,
\bea
\label{eq:2pt:xi:gen}
\langle 0 \vert \hat{\xi}_f (\vec{x},t)\hat{\xi}_g (\vec{x}',t')\vert 0\rangle &=& \int \frac{\dd^3 \vec{k}}{\left(2\pi\right)^3} \frac{\partial W(\vec{k},t)}{\partial t}\frac{\partial W(\vec{k},t')}{\partial t'} f_{\vec{k}}(t) g_{\vec{k}}^*(t') e^{-i \vec{k}\cdot(\vec{x}-\vec{x'})},
\eea 
where $f$ and $g$ can either be $\phi$ or $\pi$, and where we have used the fact that the only non-zero contribution comes from $\left<0\right|\hat{a}_{\vec{k}}\hat{a}^\dag_{\vec{q}}\left|0\right>=\delta^3(\vec{k}-\vec{q})$. In an anisotropic universe, the resulting integral is however not straightforward to compute, since both the mode functions $\phi_{\vec{k}}$ and $\pi_{\vec{k}}$, and the window function $W(\vec{k},t)$, depend on the direction of $\vec{k}$. For simplicity, we will assume that the universe is initially close to being isotropic, and that the anisotropy is only sourced by the test scalar field $\phi$. Since we want to determine the point at which the backreaction of the shear becomes problematic from the point of view of the isotropy assumption, the effect of the anisotropy of the mode and window functions can indeed be neglected, since it only gives rise to corrections of higher order in the shear. In this case, the coarse-graining scale becomes independent of $\vec{k}$, $k_\epsilon=\epsilon a \left\vert H \right\vert $. If the window function is set to a Heaviside function,
\bea 
\label{eq:Window}
W(\vec{k},t)=\Theta\left[\frac{k}{k_\epsilon(t)} -1\right],
\eea
one has $\partial W(\vec{k},t)/\partial t = -\dot{k}_\epsilon \delta(k-k_\epsilon)$. If the mode functions are isotropic, $\phi_{\vec{k}} = \phi_k$ and $\pi_{\vec{k}}=\pi_k$, the integral over the angles defining the direction of $\vec{k}$ can be performed in \Eq{eq:2pt:xi:gen}, and the resulting integral over the norm of $\vec{k}$ can be carried out with the delta function appearing in $\partial W(\vec{k},t)/\partial t$. This gives rise to
\bea
\label{eq:noise:correlator}
\langle 0 \vert \hat{\xi}_f (\vec{x},t)\hat{\xi}_g (\vec{x}',t')\vert 0\rangle &=& \underbrace{\frac{\dot{k}_\epsilon(t)}{2\pi^2} k^2_\epsilon(t) f_{k_\epsilon(t)}(t) g^*_{k_\epsilon(t)}(t)\frac{\sin\left[k_\epsilon(t) \vert \vec{x}-\vec{x}'\vert \right]}{k_\epsilon(t) \vert \vec{x}-\vec{x}'\vert}}_{(\Sigma_t)_{f,g}\left(\vert \vec{x}-\vec{x}'\vert\right)} \delta(t-t')\, ,
\eea
which shows that noises are white (\ie uncorrelated in time) in this case. This also defines the covariance matrix $\boldmathsymbol{\Sigma}$ of the stochastic noises $\xi_\phi$ and $\xi_\pi$ in the Langevin equations~\eqref{eq:Langevin:phi} and~\eqref{eq:Langevin:pi}, which can be formally integrated as
\bea
\label{eq:Phi:Solution:Green}
\bar{\boldmathsymbol{\Phi}}\left(t\right)
= \boldmathsymbol{G}(t,t_\uin)
\bar{\boldmathsymbol{\Phi}}\left(t_\uin\right) 
+ \int_{t_\uin}^t \dd s \ \boldmathsymbol{G}\left(t,s\right)
\boldmathsymbol{\xi}(s)
\, .
\eea
In this expression, we have arranged the coarse-grained phase-space coordinates into the vector $\bar{\boldmathsymbol{\Phi}} = \left(\bar{\phi}, \bar{\pi} \right)^\mathrm{T}$, and, similarly $\boldmathsymbol{\xi} = \left( \xi_\phi, \xi_\pi\right)^\mathrm{T}$. The Green's matrix $\boldmathsymbol{G}$ is associated with the homogeneous dynamics, \ie it is the solution of 
\bea
\label{eq:Green:def}
\frac{\partial}{\partial t}\boldmathsymbol{G} (t,t')=
\left(
\begin{array}{cc}
0 & a^{-3}\\
-m^2 a^3 & 0
\end{array}
\right)
\boldmathsymbol{G} (t,t') + \boldmathsymbol{I} \delta(t-t')
\eea
(see Appendix A of \Refc{Grain:2017dqa} for more details). From these relations, all statistical moments of the stochastic phase-space variables $\bar{\phi}$ and $\bar{\pi}$ can be computed. For instance, the two-point functions are given by
\bea
\label{eq:TwoPoint:phi}
\boldmathsymbol{\Xi}_{t,t'}\left(\vert \vec{x}'-\vec{x}\vert\right) \equiv 
\left\langle 0 \left\vert
\delta \bar{\boldmathsymbol{\Phi}}\left(\vec{x},t\right)
\delta \bar{\boldmathsymbol{\Phi}}\left(\vec{x}',t'\right)^\dagger
\right\vert 0 \right\rangle
= \int_{t_\uin}^{\mathrm{min}(t,t')}\dd s\,
\boldmathsymbol{G}(t,s)
\boldmathsymbol{\Sigma}_s\left(\vert \vec{x}-\vec{x}'\vert \right) 
\boldmathsymbol{G}^\dagger(t',s)
\nonumber \\
\eea
where we have defined $\delta \bar{f} = \bar{f}-\langle 0 \vert \bar{f} \vert 0\rangle$.
\subsection{Stochastic anisotropic stress}
Let us now derive the expectation value of the anisotropic stress. As announced in \Sec{sec:FormulationProblem}, since the scalar field $\phi$ is placed in a quantum state that is statistically isotropic, we will find that the mean anisotropic stress vanishes. The following calculation is however more than a simple consistency check, since it provides the building blocks for the calculation of the mean shear, which will be performed in \Sec{sec:Stochastic:Shear}. 

As can be seen in \Eq{eq:AnisotropicStress:ScalarField}, the calculation of the mean anisotropic stress requires to evaluate quantities of the type $\langle 0 \vert \partial_i \bar{\phi} \partial_j \bar{\phi} \vert 0 \rangle$. This gradient correlator can be obtained from the field correlator~\eqref{eq:TwoPoint:phi}, by expanding $\partial_i \bar{\phi} =\lim_{\alpha\to 0} [\bar{\phi}(\vec{x}+\alpha \vec{e}_i,t)-\bar{\phi}(\vec{x},t)]/\alpha$, where $\vec{e}_i$ is a unit vector pointing towards the $i^{\mathrm{th}}$ spatial direction. This allows one to write the gradient correlator as the sum of four field correlators, and in the coincident configuration where the gradients are evaluated at the same location $\vec{x}=\vec{x}'$, one finds\footnote{In the non-coincident configuration, if one expands
\bea
\vert \vec{x}+\alpha \vec{e} \vert = \vert \vec{x}\vert + \alpha \frac{\vec{x}\cdot\vec{e}}{\vert\vec{x}\vert}+\frac{\alpha^2}{2\vert\vec{x}\vert}\left[1-\left(\frac{\vec{x}\cdot\vec{e}}{\vert\vec{x}\vert}\right)^2\right]+\order{\alpha^3},
\eea
one finds
\bea
\boldmathsymbol{\Xi}_{t,t';(i,j)}(\vec{x},\vec{x}') &=& \left\langle 0 \left\vert \partial_i\bar{\boldmathsymbol{\Phi}}\left(\vec{x},t\right) \partial_j\bar{\boldmathsymbol{\Phi}}^\dagger\left(\vec{x}',t'\right) \right\vert 0 \right\rangle \nonumber \\
&=& - \frac{\boldmathsymbol{\Xi}^\prime_{t,t'}\left(\vert \vec{x}'-\vec{x}\vert\right)}{2\vert \vec{x}'-\vec{x}\vert}
-\frac{\left[\left(\vec{x}'-\vec{x}\right)\cdot \vec{e}_i\right]\left[\left(\vec{x}'-\vec{x}\right)\cdot \vec{e}_j\right]}{\vert \vec{x}'-\vec{x}\vert^2}\left[\boldmathsymbol{\Xi}^{\prime\prime}_{t,t'}\left(\vert \vec{x}'-\vec{x}\vert\right)+\frac{\boldmathsymbol{\Xi}^\prime_{t,t'}\left(\vert \vec{x}'-\vec{x}\vert\right)}{\vert \vec{x}'-\vec{x}\vert}\right]\, ,
\nonumber \\
\eea
where $\boldmathsymbol{\Xi}^\prime_{t,t'}(\vert \vec{x}'-\vec{x}\vert)$ and $\boldmathsymbol{\Xi}^{\prime\prime}_{t,t'}(\vert \vec{x}'-\vec{x}\vert)$ can be made explicit making use of \Eqs{eq:noise:correlator} and~\eqref{eq:TwoPoint:phi}. However, the coincident limit of the resulting expression is singular since $\lim_{\vert\Delta\vec{x}\vert\to 0} \Delta x_i \Delta x_j/\vert\Delta \vec{x}\vert^2$ is ill-defined (it depends on the direction of $\Delta\vec{x} \equiv \vec{x}'-\vec{x}$ along which this limit is taken), which explains why we treat the coincident configuration separately.}
\bea
\boldmathsymbol{\Xi} _{t,t';(i,j)}&\equiv& \left\langle 0 \left\vert \partial_i\bar{\boldmathsymbol{\Phi}}\left(\vec{x},t\right) \partial_j\bar{\boldmathsymbol{\Phi}}^\dagger\left(\vec{x},t'\right) \right\vert 0 \right\rangle
\nonumber \\ &=& 
 \lim_{\alpha\to 0,\beta\to 0} \frac{1}{\alpha\beta} \left[  \boldmathsymbol{\Xi}_{t,t'}\left(\left\vert \alpha \vec{e}_i-\beta\vec{e}_j\right\vert \right) +  \boldmathsymbol{\Xi}_{t,t'}\left(0\right)-\boldmathsymbol{\Xi}_{t,t'}\left(\alpha\right)-\boldmathsymbol{\Xi}_{t,t'}\left(\beta\right) \right]
\nonumber \\ & = &
 = - \boldmathsymbol{\Xi}_{t,t'}''(0) \delta_{ij}\, .
\label{eq:Xi_ij:Xi''}
\eea
In the first equality, we have used that $\boldmathsymbol{\Xi}_{t,t'}$ only depends on $\vert \vec{x}'-\vec{x}\vert$, see \Eq{eq:TwoPoint:phi}, and the primes in our final expression denote derivation with respect to that quantity. In the second equality, we have used that $\boldmathsymbol{\Xi}_{t,t'}$ is an even function of $\vert \vec{x}'-\vec{x}\vert$ (since the cardinal sine function $\mathrm{sinc}(x)\equiv\sin (x)/x$ appearing in \Eq{eq:noise:correlator} is even), hence $\boldmathsymbol{\Xi}_{t,t'}'(0)=0$; together with the fact that $\vert \alpha \vec{e}_i-\beta\vec{e}_j \vert^2 = \alpha^2+\beta^2-2\alpha\beta\delta_{ij}$. This expression can be made further explicit by making use of \Eqs{eq:noise:correlator} and~\eqref{eq:TwoPoint:phi}, and since $\mathrm{sinc}''(0)=-1/3$, one finds
\bea
\label{eq:gradient:two:point:function}
\boldmathsymbol{\Xi}_{t,t';(i,j)}=
\int_{t_\uin}^{\mathrm{min}(t,t')}\dd s\,
k_{\epsilon}^2(s)
\boldmathsymbol{G}(t,s)
\boldmathsymbol{\Sigma}_s\left(0 \right) 
\boldmathsymbol{G}^\dagger(t',s)
\frac{\delta_{ij}}{3}\, .
\eea
It is then easy to see from \Eq{eq:AnisotropicStress:ScalarField} that, in an isotropic background, the fact that $\boldmathsymbol{\Xi}^{(i,j)} _{t,t'}$ is proportional to $\delta_{ij}$ implies that
\bea
\label{eq:anisotropic:stress:one:point:function}
\left\langle 0 \left\vert  \overline{\Pi}_{ij} \right\vert 0 \right\rangle = 0\, .
\eea
This confirms that the one-point function of the anisotropic stress vanishes in an isotropic background.
\subsection{Stochastic shear}
\label{sec:Stochastic:Shear}
From \Eq{eq:AnisotropicStress:ScalarField}, the two-point function of the anisotropic stress can be expressed as combinations of four-point functions of field gradients. Since the statistics of $\phi$ is Gaussian [given that $\boldmathsymbol{\xi}$ is a Gaussian noise and that $\bar{\boldmathsymbol{\Phi}}$ is linear in the noise, see \Eq{eq:Phi:Solution:Green}], these four-point functions can be evaluated using Wick theorem, namely $\langle 0 \vert \partial_i \bar{\phi} \partial_j \bar{\phi} \partial_m \bar{\phi}\partial_n \bar{\phi} \vert 0 \rangle = \langle 0 \vert \partial_i \bar{\phi} \partial_j \bar{\phi} \vert 0\rangle \langle 0 \vert \partial_m \bar{\phi}\partial_n \bar{\phi} \vert 0 \rangle +\langle 0 \vert \partial_i \bar{\phi} \partial_m \bar{\phi}  \vert 0 \rangle \langle 0 \vert \partial_j \bar{\phi}\partial_n \bar{\phi} \vert 0 \rangle +  \langle 0 \vert \partial_i \bar{\phi} \partial_n \bar{\phi} \vert 0 \rangle \langle 0 \vert \partial_j \bar{\phi}  \partial_m \bar{\phi} \vert 0 \rangle$. This gives rise to
\bea
\left\langle 0 \left\vert \overline{\Pi}_{i}^{j} (t) \overline{\Pi}^{i}_{j} (t') \right\vert 0 \right\rangle &=&
g^{ij}(t)g^{mn}(t') \left[
{\Xi}_{t,t;(i,m)}^{\bar{\phi},\bar{\phi}} {\Xi}_{t',t';(j,n)}^{\bar{\phi},\bar{\phi}}
+{\Xi}_{t,t';(i,j)}^{\bar{\phi},\bar{\phi}} {\Xi}_{t,t';(m,n)}^{\bar{\phi},\bar{\phi}}
\right. \nonumber\\  & & \kern-2.5em + \left.
{\Xi}_{t,t';(i,n)}^{\bar{\phi},\bar{\phi}} {\Xi}_{t,t';(m,j)}^{\bar{\phi},\bar{\phi}}
-\frac{1}{3}{\Xi}_{t,t;(i,j)}^{\bar{\phi},\bar{\phi}} {\Xi}_{t',t';(m,n)}^{\bar{\phi},\bar{\phi}}
-\frac{2}{3}{\Xi}_{t,t';(i,m)}^{\bar{\phi},\bar{\phi}} {\Xi}_{t,t';(j,n)}^{\bar{\phi},\bar{\phi}}
\right].
\eea
In an isotropic universe, making use of \Eq{eq:Xi_ij:Xi''}, the terms involving products of the form ${\Xi}_{t,t;(i,j)}^{\bar{\phi},\bar{\phi}} {\Xi}_{t,'t';(k,\ell)}^{\bar{\phi},\bar{\phi}} $ cancel out. Such terms would indeed correspond to the squared expectation value of the anisotropic stress, which was already shown to vanish, see \Eq{eq:anisotropic:stress:one:point:function}. Thus only the cross terms remain, \ie those of the form ${\Xi}_{t,t';(i,j)}^{\bar{\phi},\bar{\phi}} {\Xi}_{t,t';(k,\ell)}^{\bar{\phi},\bar{\phi}} $, leading to\footnote{Here this result has been obtained by first coarse-graining the scalar field, and then calculating its two-point function and associated anisotropic stress. We have checked that by first deriving the full anisotropic stress, which includes all scales, and then coarse-graining the result, one obtains the same formula.
}
\bea
\left\langle 0 \left\vert \overline{\Pi}_{i}^{j} (t) \overline{\Pi}^{i}_{j} (t') \right\vert 0 \right\rangle = \frac{10}{a^2\left(t\right) a^2\left(t'\right)}\left[ \left(\Xi_{t,t'}^{\bar{\phi},\bar{\phi}}\right)''(0)\right]^2\, .
\label{eq:Piji:Piij:Xi''}
\eea
This can be cast in a form similar to \Eq{eq:gradient:two:point:function}, and this shows that, even in an isotropic background, scalar quantum fluctuations make the two-point function of the anisotropic stress not vanish.
\section{Shear backreaction in a contracting cosmology}
\label{sec:Contraction}
Let us now apply the calculational program sketched above to the case of a contracting cosmology (expanding cosmologies, in particular inflating cosmologies, are discussed in \Sec{sec:Discussion}). In an isotropic universe, defining $u=a \phi$, \Eqs{eq:eom:phi}-\eqref{eq:eom:pi} give rise to
\bea
\label{eq:mode:MS:u}
{u}_k''+\left(a^2m^2+k^2-\mathcal{H}^2-\mathcal{H}'\right)u_k=0\, ,
\eea
where a prime denotes derivation with respect to conformal time $\eta$, related to cosmic time through $\dd t = a \dd \eta$, and where $\mathcal{H}=a'/a$. In order to set initial conditions in the Bunch-Davies vacuum state~\cite{Bunch:1978yq}, the pulsation $\omega_k^2$, defined as the term between the braces in \Eq{eq:mode:MS:u}, must be dominated by $k^2$ at early time. In a contracting cosmology, this clearly imposes $m=0$, and for simplicity, we consider that this is the case hereafter. If the equation-of-state parameter of the background is constant, one has $\omega_k^2 = k^2-2(1-3w)/[(1+3w)^2\eta^2]$. The resulting mode equation can be solved by means of Hankel functions, and setting integration constants such that the mode functions are normalised according to the Klein-Gordon product mentioned above, one obtains
\bea
\phi_k = \frac{1}{2a}\sqrt{\frac{\pi}{k}} e^{i\frac{\pi}{4}(2\nu+1)} \sqrt{-k \eta}H_{\nu}^{(1)}\left(-k\eta\right) 
\quad\quad \mathrm{with}\quad\quad \nu=\frac{3}{2} \frac{1-w}{1+3w}  
\eea
in the vacuum state, where $H_{\nu}^{(1)}$ is the first Hankel function and where, hereafter, we assume that $-1/3<w<1$. On super-Hubble scales, the Hankel function can be expanded, and one finds $\phi_{k\ll a H} \simeq  \ee^{i\pi (2\nu-1)/4}2^{\nu-1}\Gamma(\nu)(-k\eta)^{1/2-\nu}/(\sqrt{\pi k}a)$ and $\pi_k = a^2\phi_k'$. Plugging these expressions into \Eq{eq:noise:correlator}, one obtains
\bea
\kern-2em
\boldmathsymbol{\Sigma}_t\left(\vert \vec{x}-\vec{x}'\vert\right) &=& \displaystyle \frac{\Gamma^2(\nu)}{\left(2\pi\right)^3 }
\left(1+3w\right)^{\frac{3(1-w)}{1+3w}} \epsilon^{\frac{12w}{1+3w}}\left\vert H\right\vert^3
\frac{\sin\left[k_\epsilon(t) \vert \vec{x}-\vec{x}'\vert \right]}{k_\epsilon(t) \vert \vec{x}-\vec{x}'\vert}
\nonumber \\  &\quad & \times
 \left(
\begin{array}{ccc}
1 &\quad\quad & \frac{3}{2}\left(w-1\right) a^3 H\\
\frac{3}{2}\left(w-1\right) a^3 H & \quad\quad& \left[\frac{3}{2}\left(w-1\right) a^3 H\right]^2
\end{array}
\right)\, .
\nonumber\\
\label{eq:Sigma:MasslessField}
\eea

If the mass vanishes, the background equation of motion~\eqref{eq:Green:def} can be readily solved, and one obtains for the Green function
\bea
\boldmathsymbol{G} (t,t') = 
\left( \begin{array}{cc}
1& \int_{t'}^t \frac{\dd s}{a^3(s)}\\
0 & 1
 \end{array}\right)\Theta\left(t-t'\right)\, .
 \label{eq:Green:massless}
\eea
Recalling that $\mathrm{sinc}''(0)=-1/3$, by making use of \Eqs{eq:noise:correlator} and~\eqref{eq:TwoPoint:phi}, \Eq{eq:Piji:Piij:Xi''} then gives rise to
\bea
\kern-2em
\left\langle 0 \left\vert \overline{\Pi}_{i}^{j} (t_1) \overline{\Pi}^{i}_{j} (t_2) \right\vert 0 \right\rangle &= & \frac{5 \Gamma^4\left(\frac{3}{2} \frac{1-w}{1+3w}  \right)}{288\pi^6} \frac{\left(1+3w\right)^{\frac{6(1-w)}{1+3w}}}{\left(1+9w\right)^2}\epsilon^{\frac{4(9w+1)}{1+3w}}H_\uin^{8} \left(\frac{a_1a_2}{a_\uin^2}\right)^{{3w-5}}
\nonumber \\ & & \kern2em
\left\lbrace  \left[\mathrm{\max} \left(\frac{a_1}{a_\uin},\frac{a_2}{a_\uin}\right)\right]^{-1-9w}-1\right\rbrace^2  ,
\label{eq:Piji:Piij:Xi'':explicit}
\eea
where $a_1$ and $a_2$ denote $a(t_1)$ and $a(t_2)$ respectively. The contribution from the shear to the overall energy density can then be computed with \Eq{eq:rhoSigma:AnisotropicStress}. In order to specify the term $\mathrm{max}(a_1,a_2)$ appearing in \Eq{eq:Piji:Piij:Xi'':explicit}, the rectangular integration domain of \Eq{eq:rhoSigma:AnisotropicStress} can be split into two triangles, one where $t_1<t_2$ and one where $t_1>t_2$. Since the integrand is in fact symmetric in $t_1$ and $t_2$, both these triangular regions give the same contribution and it is enough to compute the integral over one of the triangles and multiply it by two. This leads to
\bea
\left\langle 0 \vert  \rho_\sigma \vert 0 \right\rangle& =& \frac{5}{432\pi^6}
\Gamma^4\left(\frac{3}{2}\frac{1-w}{1+3w}\right)
\frac{\left(1+3w\right)^9}{\left(1+9w\right)^2}\left(\frac{\epsilon}{1+3w}\right)^{4\frac{1+9w}{1+3w}} \frac{H^6}{\Mp^2} \left[
\frac{1}{5+27 w}
-\left(\frac{a}{a_\uin}\right)^{1+9w}
\right. \nonumber \\  & & \kern-3em \left.
+\frac{3+9w}{\left(1-9w\right)^2}\left(\frac{a}{a_\uin}\right)^{2+18w}
-\frac{16\left(1+9w\right)^2}{\left(1-9w\right)^2\left(5+27w\right)}\left(\frac{a}{a_\uin}\right)^{\frac{5+27w}{2}}
+\left(\frac{1+9w}{1-9w}\right)^2\left(\frac{a}{a_\uin}\right)^{3+9w}
\right] .
\nonumber \\
\label{eq:rho_shear:contraction}
\eea
One can check that, when $t=t_\uin$, $\left\langle \rho_\sigma \right\rangle=0$. When $w>-1/9$, the terms have been ordered in decreasing power, while for $w<-1/9$, the dominant term is the third one (\ie the first one on the second line), which increases as $\propto a^{2+18w}$. These two cases need therefore to be distinguished. 

\begin{figure}[t]
\begin{center}
\includegraphics[width=0.496\textwidth]{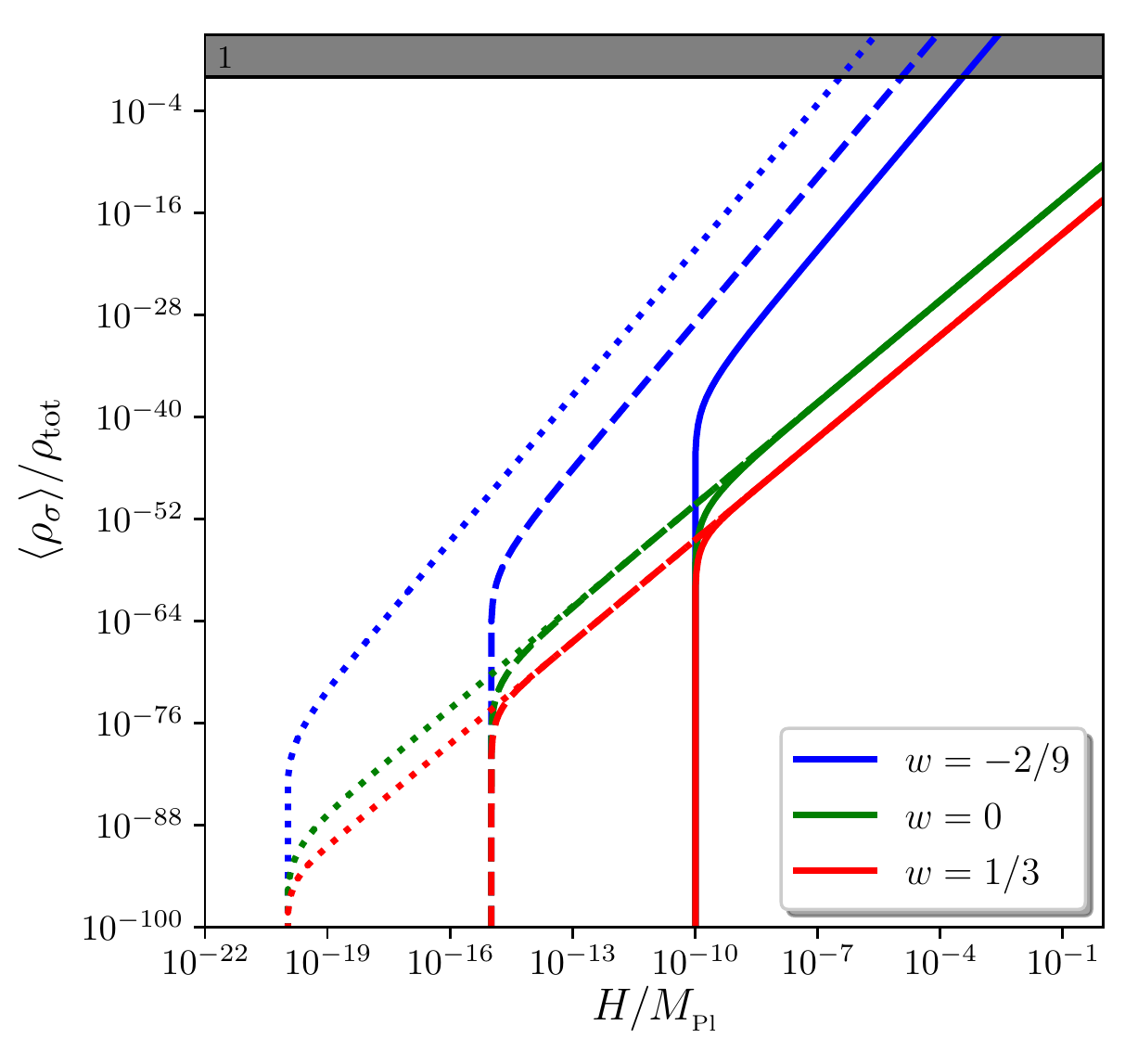}
\includegraphics[width=0.496\textwidth]{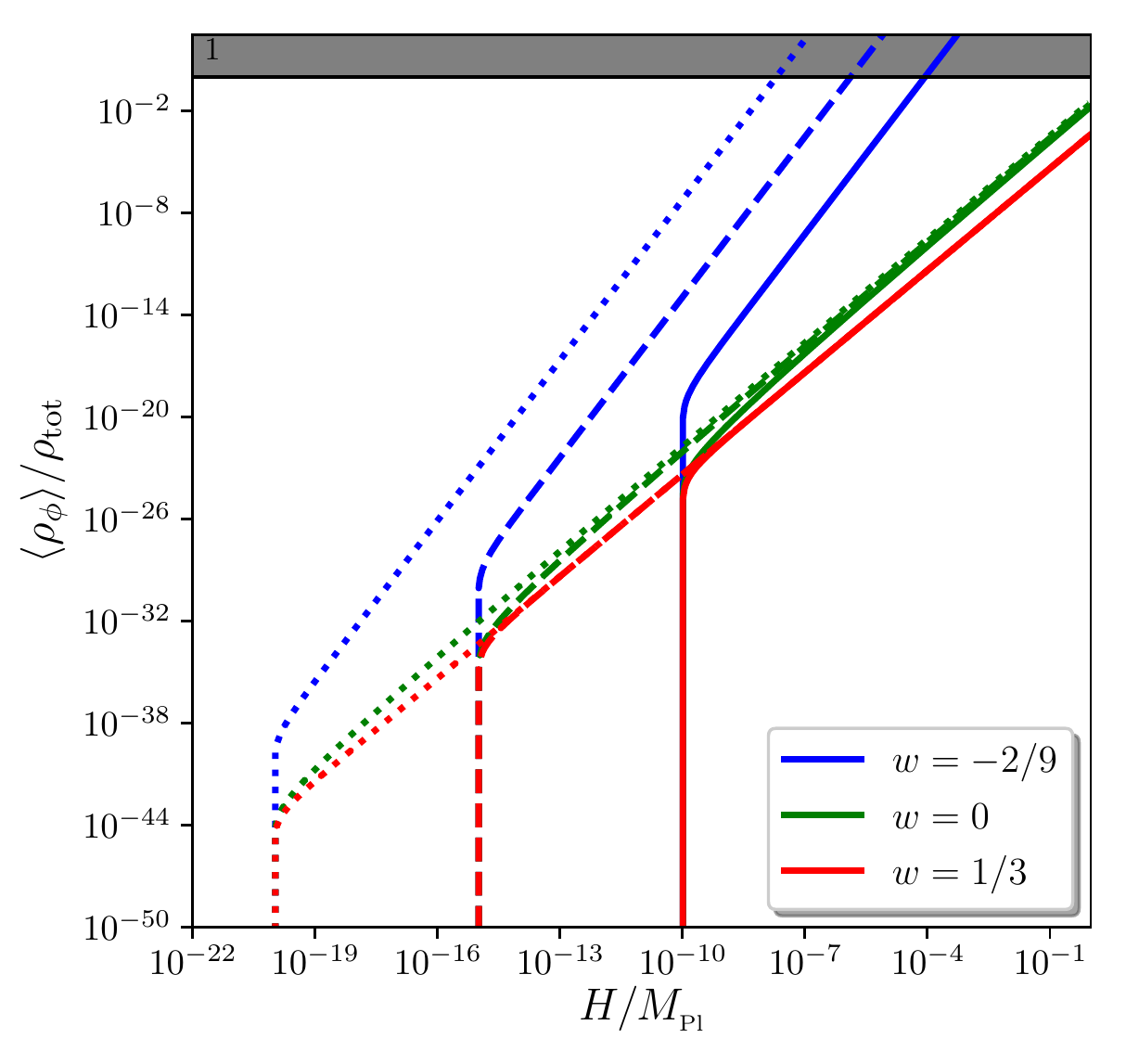}
\caption{Left panel: contribution from the shear to the overall energy density in a contracting cosmology, as a function of time labeled by the value of the Hubble scale in Planckian units. The result is computed with \Eq{eq:rho_shear:contraction}, using $\epsilon=0.1$ Three different equation-of-state parameters are displayed, namely $w=-2/9$ (blue lines), $w=0$ (green lines) and $w=1/3$ (red lines), for three different initial times, $H_\uin/\Mp = 10^{-10}$ (solid lines), $H_\uin/\Mp = 10^{-15}$ (dashed lines) and $H_\uin/\Mp = 10^{-10}$ (dotted lines). The grey shaded area corresponds to $\langle\rho_\sigma\rangle>\rho_{\mathrm{tot}}$, where the backreaction from the shear becomes important (and where our calculation does not apply). Right panel: direct contribution from the scalar field fluctuations to the overall energy density, in the same situations as those shown in the left panel.}  
\label{fig:shear}
\end{center}
\end{figure}
When $w>-1/9$, since the dominant term, \ie the one proportional to $1/(5+27 w)$, does not depend on the initial time (contrary to the other terms, it does not involve $a_\uin$), the contribution from the shear to the energy density quickly reaches an attractor, which scales as $\langle \rho_\sigma\rangle \propto H^6/\Mp^2$, hence $\langle \rho_\sigma\rangle /\rho_{\mathrm{tot}}  \propto (H/\Mp)^4$. As a consequence, until the background energy density reaches the Planck scale, the contribution from the shear to that energy density is negligible.  This can be clearly seen in the left panel of \Fig{fig:shear} (see the cases $w=0$ and $w=1/3$). In fact, because of the prefactor in \Eq{eq:rho_shear:contraction}, the ratio $\langle \rho_\sigma\rangle /\rho_{\mathrm{tot}}$ is still negligible when $H$ reaches the Planck scale. In bouncing cosmologies, $|H|$ is bounded from above since the bounce demands for a smooth transition from contraction, \ie $H<0$, to expansion, \ie $H>0$. The upper-bound of $|H|$ can be safely taken to be of the order of the Planck scale at most\footnote{Either the bounce is classical and we can expect the maximal value of $|H|$ to be below the Planck scale; or the bounce is of quantum-gravitational origin, hence expected to occur at the Planck energy density, as can explicitly be proven in Loop Quantum Cosmology for example.} and the maximal value that $\langle \rho_\sigma\rangle /\rho_{\mathrm{tot}}$ can achieve is thus of order $\sim\epsilon^{4\frac{1+9w}{1+3w}}$. Hence, one concludes that there is no shear backreaction problem in this case.  

When $w<-1/9$, since the dominant term, \ie the first one of the second line in \Eq{eq:rho_shear:contraction}, explicitly depends on $a_\uin$, the late-time behaviour of $\langle\rho_\sigma\rangle$ strongly depends on the initial time of the contraction and there is no attractor. Besides, this dominant term increases in time (recall that $a$ decreases in a contracting cosmology), so $\langle \rho_\sigma\rangle /\rho_{\mathrm{tot}}$ does not scale as $(H/\Mp)^4$ anymore, but rather as $(H/\Mp)^{2(7-9w)/[3(1+w)]} [\Mp/(\epsilon H_\uin)]^{-4(1+9w)/[3(1+w)]}$. 
This implies that, if the contracting phase starts in the infinite past ($\epsilon H_\uin=0$), then there is an IR divergence in the shear. Indeed, if $w<-1/9$, then the power spectrum of the field fluctuations is too red for the large wavelengths to give a finite contribution to the shear. This problem can be solved by considering a finite contraction. However, if $H_\uin$ is sufficiently small, \ie if the contracting phase starts sufficiently early on, the contribution from the shear to the overall energy density becomes sizeable before the energy density of the universe reaches the Planck scale. This is again well visible in the left panel of \Fig{fig:shear}, see the case $w=-2/9$. Whether or not a backreaction problem occurs before $H$ reaches $\Mp$ thus depends on the value of $H_\uin$, \ie on the total amount of contraction. This is why, in \Fig{fig:HinMin}, we display the minimum value of $H_\uin$ such that the shear does not backreact on the background geometry until $H$ reaches the Planck scale. One can see that, when $w\lesssim -0.2$, this is impossible to achieve.

Let us note that the case $w=-1/9$ is a priori singular. Indeed, in the prefactor of  \Eq{eq:rho_shear:contraction}, one notices the presence of the term $(1+9w)^{-2}$, which diverges when $w$ approaches $-1/9$. However, in that limit, all three first terms in the square brackets become of the same order, and, when combined together, they precisely give a contribution proportional to $(1+9w)^{2}$ (with a logarithmic dependence on $a/a_\uin$), making this limit regular.

In all cases, let us also stress that quantum diffusion makes the shear grow faster than classically. Indeed, if $w>-1/9$, one has $\langle\rho_\sigma\rangle \propto a^{-9(1+w)}$ and if $w<-1/9$, $\langle\rho_\sigma\rangle \propto a^{9w-7}$. So $\langle\rho_\sigma\rangle$ always increases faster than $a^{-8}$, while it scales as $a^{-6}$ classically. 
\section{Discussion}
\label{sec:Discussion}
\begin{figure}[t]
\begin{center}
\includegraphics[width=0.496\textwidth]{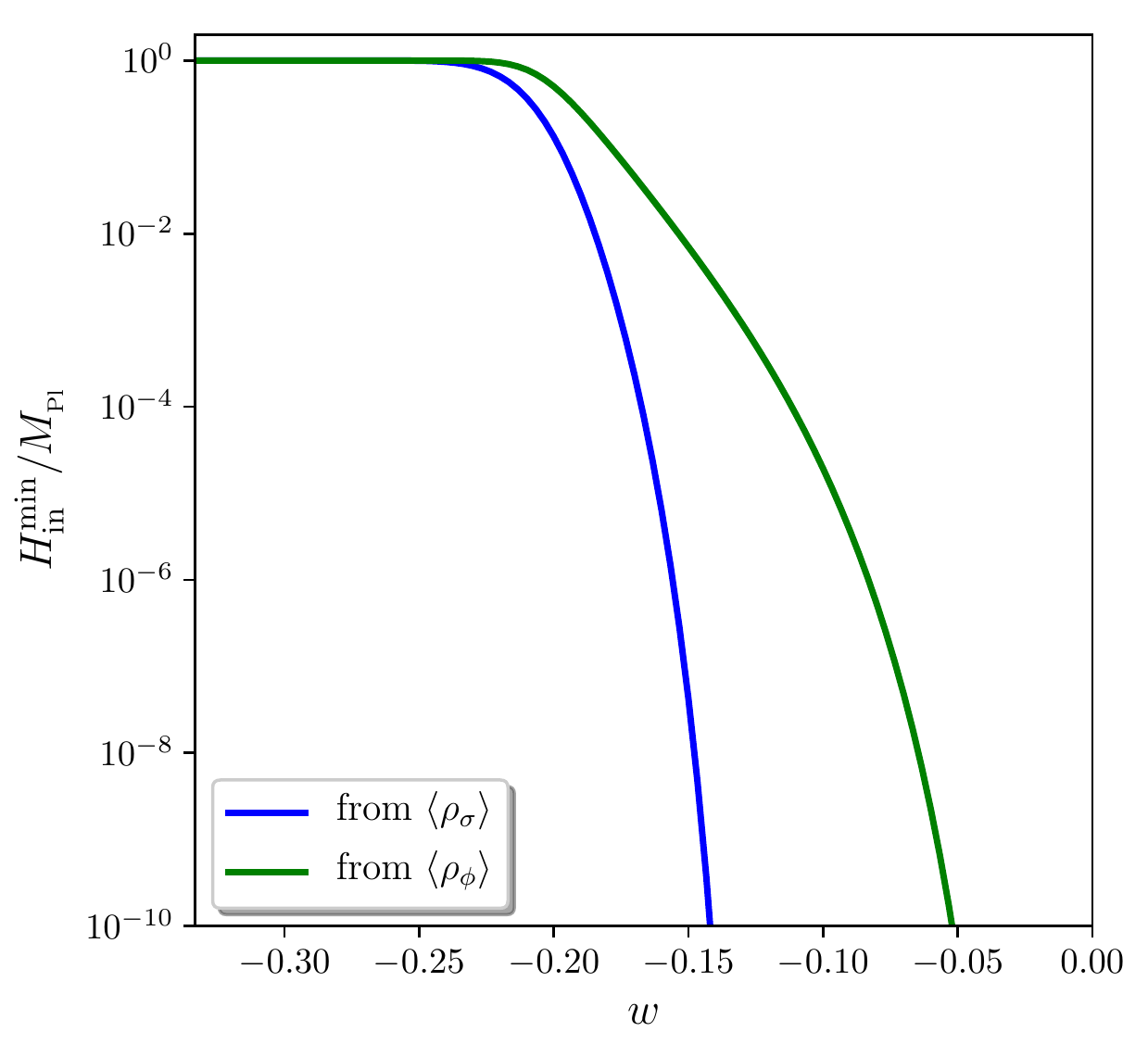}
\caption{Minimum value of the initial Hubble parameter (parametrising the maximum amount of contraction) such that the backreaction coming from the shear (blue curve) and from the spectator field fluctuations (green) remain negligible until the time when the overall energy density reaches the Planck scale.}  
\label{fig:HinMin}
\end{center}
\end{figure}
For comparison, it is interesting to compute the contribution to the overall energy density  arising directly from the scalar field fluctuations themselves. Indeed, from \Eq{eq:StressEnergyTensor:ScalarField}, the ``$00$'' component of the stress-energy tensor reads $\rho_\phi=\dot{\phi}^2/2+V(\phi)+\delta^{ij}\partial_i\phi\partial_j\phi/(2a^2)$. In the absence of a potential, and after neglecting gradient terms, which are suppressed by higher powers of $\epsilon$ in the coarse-grained sector, this gives rise to\footnote{In this case, by computing the ``$(ij)$'' component of the stress-energy tensor in \Eq{eq:StressEnergyTensor:ScalarField}, one can see that the pressure associated with $\phi$ takes the same form, \ie $p_\phi=\rho_\phi$. This implies that the direct fluctuations of the scalar field also have an effective equation-of-state parameter $w=1$ (provided the potential vanishes), as for the shear (for which $w=1$ even in the presence of a non-vanishing potential). At the classical level, these two therefore backreact in the same way, but at the quantum level, they behave differently, see \Eqs{eq:rho_shear:contraction} and~\eqref{eq:rho:phi:contraction}.} $\rho_\phi=\dot{\phi}^2/2=\pi^2/(2 a^6)$, see \Eq{eq:eom:phi}. One thus has to compute $\langle \rho_\phi \rangle =\Xi_{t,t}^{\bar{\pi},\bar{\pi}}(0) /(2 a^6)$, where the two-point functions $\boldmathsymbol{\Xi}_{t,t'}(\vert \vec{x}'-\vec{x}\vert)$ are given in \Eq{eq:TwoPoint:phi}. Making use of \Eq{eq:Sigma:MasslessField} for the noise correlator and of \Eq{eq:Green:massless} for the Green function, this gives rise to
\bea
\left\langle 0 \vert \rho_\phi \vert 0 \right\rangle = \frac{3\Gamma^2\left(\frac{3}{2} \frac{1-w}{1+3w}\right)}{128\pi^3}\frac{\left(1+3w\right)^3\left(w-1\right)^2}{w}\left(\frac{\epsilon}{1+3w}\right)^{\frac{6w}{1+3w}} H^4 \left[1-\left(\frac{a}{a_\uin}\right)^{6w}\right]
\label{eq:rho:phi:contraction}
\eea
if $w\neq 0$, while $\langle \rho_\phi \rangle = 9 H^4/(256\pi^2)\ln(a_\uin/a)$ when $w=0$. One can see that two regimes need again to be distinguished. If $w>0$, the second term in the square brackets, $(a/a_\uin)^{6w}$, quickly becomes negligible as contraction proceeds, and one reaches an attractor where $\langle \rho_\phi\rangle \propto H^4$, hence $\langle \rho_\phi\rangle/\rho_{\mathrm{tot}}$ is suppressed by $(H/\Mp)^2$. In this case, the spectator field does not backreact until $H$ reaches the Planck scale, and the direct contribution from the scalar field fluctuations to the overall energy density is larger than that of the shear, which we recall is suppressed by $(H/\Mp)^4$. If $w<0$, the second term in the square brackets provides the dominant contribution. Since it depends explicitly on $a_\uin$, there is no attractor anymore, $\langle \rho_\phi\rangle/\rho_{\mathrm{tot}}$ scales as $(H/\Mp)^{2(1-w)/(1+w)} (\Mp/H_\uin)^{-4w/(1+w)}$, and there is an IR divergence (similar to the one discussed for the shear) if the contracting phase extends infinitely far in the past. The energy density contained in $\phi$ can therefore become large before $H$ reaches the Planck mass, provided $H_\uin$ is small enough, \ie provided contraction lasts long enough. The direct contribution from the scalar field fluctuations to the overall energy density is displayed in the right panel of \Fig{fig:shear}, and in \Fig{fig:HinMin}, we show the minimum value of the initial Hubble rate such that no backreaction occurs, as a function of the equation-of-state parameter. Since this value is always larger than the one coming from the requirement of no shear backreaction, one concludes that the direct contribution of the scalar field fluctuations to the energy density becomes sizeable before its contribution to the shear does. As a consequence, the assumption of working with a test field breaks down before the shear becomes sizeable, and one would need to relax it in order to investigate the dynamics of the universe when it becomes truly anisotropic.

Let us note that, in an inflating cosmology, a light ($m\ll H$) spectator scalar field also generates an anisotropic stress, although there is no classical shear instability in that case. In a de-Sitter space-time, once the field has reached the slow-roll attractor, at the background level, it evolves as $\phi(t) = e^{-m^2(t-t_\uin)/(3H)}\phi(t_\uin)$, and if its mode functions are normalised to the Bunch-Davies vacuum, on super-Hubble scales, they are given by $\phi_{k\ll aH}\simeq 2^{\nu-1}\Gamma(\nu)H^{\nu-1/2}a^{\nu-3/2}/(k^\nu\sqrt{\pi})$, where $\nu=3/2\sqrt{1-4m^2/(9 H^2)}$. Using the same procedure as above, this gives rise to
\bea
\left\langle 0 \vert \rho_\sigma \vert 0\right\rangle& =& \frac{2^{4\left(\nu-2\right)} \Gamma^4\left(\nu\right)}{27 \pi^6} \frac{\epsilon^{10-4\nu}\frac{H^6}{\Mp^2}}{\left(1+\frac{m^2}{3H^2}\right)^2\left(1+\frac{2 m^2}{15 H^2}\right)}\Bigg[
1-5 \frac{1+\frac{2 m^2}{15 H^2}}{1-\frac{ m^2}{6 H^2}}\left(\frac{a_\uin}{a}\right)^{2+\frac{2m^2}{3H^2}}
\nonumber \\ & &  
+15 \frac{1+\frac{2 m^2}{15 H^2}}{\left(1-\frac{ 2m^2}{3 H^2}\right)^2}\left(\frac{a_\uin}{a}\right)^{4+\frac{4m^2}{3H^2}}
-16 \frac{\left(1+\frac{ m^2}{3 H^2}\right)^2}{\left(1-\frac{ 2m^2}{3 H^2}\right)^2}\left(\frac{a_\uin}{a}\right)^{5+\frac{2m^2}{3H^2}}
 \nonumber \\ & & 
+ 5 \frac{\left(1+\frac{ m^2}{3 H^2}\right)^2 \left(1+\frac{2m^2}{15H^2}\right)}{\left(1-\frac{ 2 m^2}{3 H^2}\right)^2 \left(1-\frac{m^2}{6H^2}\right)}\left(\frac{a_\uin}{a}\right)^{6}
\Bigg] .
\eea
One can see that, up to quickly decaying terms, $\rho_\sigma$ acquires a constant value, of order $H^6/\Mp^2$. Therefore, its relative contribution to the Friedmann equation~\eqref{eq:Friedmann} is suppressed by $(H/\Mp)^4$ and remains small at sub-Planckian scales. For instance, in the light-field limit, \ie when $m\ll H$, the above expression gives 
\bea
\frac{\langle 0 \vert \rho_\sigma \vert 0 \rangle }{3\Mp^2H^2} \simeq \left(\frac{\epsilon H}{6\pi\sqrt{2}\Mp}\right)^4.
\eea
In single-field slow-roll models of inflation, with current measurements~\cite{Aghanim:2018eyx,Akrami:2018odb} of the scalar power spectrum amplitude $\calP_\zeta = H^2/(8\pi^2\epsilon_1\Mp^2)\simeq 2\times 10^{-9}$ where $\epsilon_1$ is the first slow-roll parameter, and current constraints from the Planck data on the tensor-to-scalar ratio $r=16\epsilon_1<0.1$, one obtains $\rho_\sigma/(3\Mp^2 H^2) < 10^{-24}\epsilon^4$ at the time when observed scales are produced, a tiny value indeed. Let us note however that, compared to the Friedmann equation, the relative contribution of the shear to the Raychaudhuri equation~\eqref{eq:Raychaudhuri} is enhanced by an additional factor $1/\epsilon_1$, since $\dot{H}=-\epsilon_1 H^2$, but using the above formulas the contribution is proportional to $\epsilon^4 (H/\Mp)^2 \calP_\zeta$ and is therefore still negligible. Let us finally point out that the energy density contained in the scalar field fluctuations reads $\langle \rho_\phi \rangle \simeq m^2 \langle\bar{\phi}^2\rangle/2 \simeq \epsilon^{3-2\nu} 2^{2\nu-5} 3 \Gamma^2(\nu) H^4/\pi^3$, where one has used \Eq{eq:TwoPoint:phi} to evaluate $\langle\bar{\phi}^2\rangle$. This is in fact larger than $\langle \rho_\sigma \rangle$, so $\phi$ would stop acting as a test field before it generates a substantial shear, as in the contracting case discussed in \Sec{sec:Contraction}. There is therefore no shear backreaction problem in inflating cosmologies. This is in agreement with \Refc{DeAngelis:2020wjp}, where, instead of working with Bianchi I cosmologies, the effect of quantum anisotropies on quasi de-Sitter backgrounds was studied in the context of Taub cosmologies. In \Refc{Wolfson:2020fqz}, the classical isotropisation of inflating space times in the presence of SU(2)-gauge fields coupled with an axion was also studied, and it was shown that some initial configurations lead to a premature termination of slow-roll inflation if the axion drives inflation. In \Refc{Fujita:2017lfu}, the case of inflating Bianchi I space times in the presence of a U(1) gauge field coupled to the inflaton was also investigated, and the stochastic formalism was employed to show that the quantum fluctuations in the gauge field can produce a large statistical anisotropy.
\section{Conclusion}
\label{sec:Conclusion}
Isotropic contracting cosmologies with $w<1$ are unstable fixed points, since any amount of initial shear leads to an effective contribution to the energy density that grows as $a^{-6}$. Nonetheless, if one sets initial conditions in an exactly isotropic universe, classically, the universe remains isotropic. When quantum fluctuations are included however, they make the system diffuse away from the isotropic fixed point, and the classical instability then has the potential to drive a shear backreaction problem. In this work, we have quantified this effect if the anisotropic stress arises from quantum fluctuations of a test, massless scalar field. We have found that, if $w>-1/9$, the shear contribution to the overall energy density reaches an attractor that remains negligible until the energy density of the universe reaches the Planck scale, where a bounce is expected to take place (and close to which our treatment does not apply anyway). If $w<-1/9$ however, the shear backreaction can become substantial before one reaches the Planck scale, if the contracting phase lasts long enough (which signals the presence of an IR divergence for an infinite contraction). In either case, we have also found that the direct contribution from the scalar field quantum fluctuations to the energy density of the universe is always more important than that coming from the shear. As a consequence, in order to study the dynamics of the universe as it becomes truly anisotropic, one would first need to incorporate the backreaction of the scalar field on the overall energy density. 

It is worth stressing that by comparing $\langle \rho_\sigma \rangle$ to the energy density associated to the background, we have tried to estimate the probability that a substantial shear is imprinted on a given FLRW patch. Our result is therefore stochastic by nature, \ie $\rho_\sigma$ should not be thought of as an actual energy density component but more as an indicator for the presence of a potential backreaction problem. This also indicates that one may gain further insight by computing the full probability distribution function associated to the shear. Since it is related to the scalar field fluctuations in a non-linear way, its statistics should be non-Gaussian, and we plan to further investigate it in future works.

Let us also note that the stochastic formalism was already employed in the context of contracting cosmologies in \Refc{Miranda:2019ara}, where the case of a contracting phase driven by a scalar field with an exponential potential was studied. In such systems, phase space possesses classical unstable fixed points. Even if one sets initial conditions exactly on such fixed points, quantum fluctuations drive the system away from these configurations, and the outcome of this quantum instability was found to depend on the equation-of-state parameter during the contraction: if $w=1/3$, there is no substantial backreaction before one reaches the Planck scale, while if $w=0$, the quantum instability can become effective before the Planck scale if contraction lasts long enough. Those results are therefore very similar to ours. 

Although our calculation was performed for a massless spectator scalar field, it can be easily generalised to other setups. For instance, if the mass of the field is not vanishing, the dynamics remains linear and the methods presented in this work can still be directly applied. However, in that case, three physical scales play a role in the dynamics of a given Fourier mode: its physical wavenumber $k/a$, the Hubble scale $H$, and the mass of the field $m$. In the asymptotic past, both $k/a$ and $H$ become small, but $k/a$ takes over $H$ if $w>-1/3$ and the effect of the expansion becomes negligible. However, since $m$ is constant, it provides the dominant contribution to the frequency of the field fluctuations in the asymptotic past. Since they cannot be described as massless fluctuations on a flat space-time in that limit, the Bunch-Davies vacuum cannot be imposed, which makes the choice of the initial quantum state more involved. 

One may however speculate that, as long as the field is heavy, it remains unexcited, and that it starts sourcing the shear only when $H$ becomes larger than $m$. From that point on, since the field is light, our results may apply, which means that, in practice, one might simply have to set the initial value of the Hubble parameter $H_\uin$ to $m$. If this is correct, the bound we have obtained on $H_\uin$ for soft equations of state, see the discussion around \Fig{fig:HinMin}, translates into a lower bound on $m$, which is of the order of the Planck mass. The conclusion would therefore be that fields with masses substantially below the Planck mass make isotropic contracting cosmologies with soft equations of state unstable. However, one would have to check that quickly after the time when the field becomes light, sub-Hubble fluctuations relax to the Bunch-Davies state, which is a priori not guaranteed (this is similar to the trans-Planckian problem~\cite{Martin:2000xs} in the inflationary context). We plan to address these questions in the future.

The geometry in which the anisotropic degrees of freedom are allowed to develop could finally be extended to more generic setups than Bianchi I metrics. The case of non-test fields could also be studied. When the contribution of the anisotropic stress to the background dynamics becomes sizeable, its backreaction could be modelled with the stochastic formalism (provided the separate-universe approach holds in that case), which may help to understand how the geometry of the universe behaves as one approaches the bounce. This would notably imply to compute the field perturbations on an anisotropic configuration, and to incorporate anisotropic noises in the Langevin equations. We leave these extensions for future work.

\begin{acknowledgments}
It is a pleasure to thank Emmanuel Frion, Tays Miranda, Patrick Peter and David Wands for very interesting comments and discussions.
\end{acknowledgments}

\appendix

\section{Comparison between stochastic and quantum-field-theoretic calculations}
\label{sec:app:QFT}

In this appendix, we explain how our results could be obtained by performing a full-quantum-field theoretic (QFT) calculation, and show that the stochastic procedure employed in this work provides a good approximation to the full QFT result.

The stochastic procedure leading to the expectation value of $\rho_\sigma$ relies on two approximations. First, by identifying the quantum operators $\hat{\xi}_\phi$ and $\hat{\xi}_\phi$, defined in \Eq{eq:xi:phi} and~\eqref{eq:xi:pi}, with their stochastic counterparts, \ie with Gaussian stochastic noises sharing the same two-point function~\eqref{eq:2pt:xi:gen}, non-vanishing commutators are neglected. However, in the present case, since $\langle  \hat{\rho}_\sigma \rangle$ involves four-point functions of $\phi$ only (\ie it does not involve $\pi$, the momentum conjugated to $\phi$), there is no effect from non-vanishing commutators in that quantity, and the stochastic procedure is exact with that respect (see \Refs{Martin:2015qta, Grain:2017dqa}). Second, once a given Fourier mode has joined the coarse-grained sector, \ie once it has crossed out the coarse-graining scale $k_\epsilon(t)$, it is evolved with the background equations of motion~\eqref{eq:Langevin:phi} and~\eqref{eq:Langevin:pi} while it should in principle be evolved with the full equations of motion~\eqref{eq:eom:phi} and~\eqref{eq:eom:pi}, where the gradient terms are not neglected. The non-inclusion of gradient terms on super coarse-graining scales leads to an error that we now try to characterise. 

We consider a free spectator scalar field and employ the same vectorial notations as in \Sec{sec:Langevin:Free:Scalar}: $\bar{\boldsymbol{\Phi}}:=\left(\bar\phi,\bar\pi\right)^\mathrm{T}$, $\boldsymbol{\xi}:=\left(\xi_\phi,\xi_\pi\right)^\mathrm{T}$ and $\boldsymbol{\Phi}_k:=\left(\phi_k,\pi_k\right)^\mathrm{T}$. The scale of coarse-graining, $k_\epsilon(t)$, is labelled by the small parameter $\epsilon$. It is a bijective function from $t$ to $k$ (\ie $k=k_\epsilon(t)$), and we denote its inverse bijection from $k$ to $t$ as $k_\epsilon^{-1}$ (\ie $t=k_\epsilon^{-1}(k)$). Following the prescription of \Sec{sec:Langevin:Free:Scalar}, we let $k_\epsilon(t) = \epsilon a(t) \vert H(t) \vert$, and use a step function for the window function, see \Eq{eq:Window}, which leads to $\partial W(\vec{k},t)/\partial t = -\dot{k}_\epsilon \delta(k-k_\epsilon)$. In vectorial notations, \Eqs{eq:xi:phi}-\eqref{eq:xi:pi} can be recast as
\bea
	\boldsymbol{\xi}=-\ds\int\frac{\dd^3\vec{k}}{(2\pi)^{3/2}}\dot{W}(t)\left[\hat{a}_ke^{-i\vec{k}\vec{x}}\boldsymbol{\Phi}_k(t)+\hat{a}^\dag_ke^{i\vec{k}\vec{x}}\boldsymbol{\Phi}^\star_k(t)\right] .
\eea
The Langevin equations~\eqref{eq:Langevin:phi} and~\eqref{eq:Langevin:pi}, and the equations of motion for the mode functions~\eqref{eq:eom:phi} and~\eqref{eq:eom:pi}, read
\bea
	\dot{\bar{\boldsymbol{\Phi}}}&=&\boldsymbol{A}_{(0)}\bar{\boldsymbol{\Phi}}+\boldsymbol\xi, \\
	\dot{\boldsymbol{\Phi}}_k&=&\boldsymbol{A}_{(k)}\boldsymbol{\Phi}_k,
\eea
with
\bea
	\boldsymbol{A}_{(k)}(t):=\left(\begin{array}{cc}
		0 & a^{-3} \\
		-a^3\left(\frac{k^2}{a^2}+m^2\right) & 0
	\end{array}\right)=\underbrace{\left(\begin{array}{cc}
		0 & a^{-3} \\
		- m^2 a^3 & 0
	\end{array}\right)}_{\boldsymbol{A}_{(0)}}+\underbrace{\left(\begin{array}{cc}
		0 & 0 \\
		-ak^2 & 0
	\end{array}\right)}_{\boldsymbol{V}_{(k)}}.
\eea
The Green function introduced in \Eq{eq:Green:def} generates the solutions of the classical background dynamics, driven by $\boldsymbol{A}_{(0)}$. For this reason, we denote it  $\boldsymbol{G}_{(0)}(t,s)$ in this appendix, while $\boldsymbol{G}_{(k)}(t,s)$ stands for the solution of the quantum mode functions, \ie is the solution of \Eq{eq:Green:def} where $\boldsymbol{A}_{(0)}$ is replaced with $\boldsymbol{A}_{(k)}$. With these notations, on the one hand, \Eqs{eq:TwoPoint:phi} and~\eqref{eq:noise:correlator} can be recast as
\bea
\label{eq:Xi:stochastic}
\boldmathsymbol{\Xi}_{t,t'}\left(\vert \vec{x}'-\vec{x}\vert\right) &=& \frac{1}{6\pi^2}
\int_{t_\uin}^{\mathrm{\min}(t,t')} \dd s  \frac{\dd k_{\epsilon}^3(s)}{\dd s} 
\mathrm{sinc}\left[k_\epsilon(s)\vert \vec{x}'-\vec{x}\vert\right]
\nonumber \\ & & \times
\boldsymbol{G}_{(0)}(t,s)\boldsymbol{\Phi}_{k_\epsilon(s)}(s)\boldsymbol{\Phi}^\dag_{k_\epsilon(s)}(s)\boldsymbol{G}^\dag_{(0)}(t',s)\, .
\eea
On the other hand, a full QFT calculation would instead lead to the following covariance matrix 
\bea
\boldmathsymbol{\Xi}_{t,t'}^\mathrm{QFT}\left(\vert \vec{x}'-\vec{x}\vert\right) &=& \frac{1}{6\pi^2}\ds\int^{k_{\epsilon}[\mathrm{min}(t,t')]}_{k_\epsilon(t_\uin)}\dd k^3
\mathrm{sinc}\left(k\vert \vec{x}'-\vec{x}\vert\right)
\boldsymbol{\Phi}_{k}(t)\boldsymbol{\Phi}^\dag_{k}(t') .
\eea
To make the comparison between these two expressions explicit, let us first note that 
\bea
	\boldsymbol{\Phi}_{k}(t)=\boldsymbol{G}_{(k)}(t,s)\boldsymbol{\Phi}_{k}(s).
\eea
In the above, the time $s$ can be arbitrarily chosen, with the only constraint that is has to be anterior to $t$. Let us thus choose $s=t_k=k^{-1}_\epsilon(k)$, which is possible thanks to the bijective nature of $k_\epsilon(t)$. This leads to
\bea
\boldmathsymbol{\Xi}_{t,t'}^\mathrm{QFT}\left(\vert \vec{x}'-\vec{x}\vert\right) &=& \frac{1}{6\pi^2}\ds\int^{k_{\epsilon}[\mathrm{min}(t,t')]}_{k_\epsilon(t_\uin)}\dd k^3
\mathrm{sinc}\left(k\vert \vec{x}'-\vec{x}\vert\right)
\boldsymbol{G}_{(k)}(t,t_k)\boldsymbol{\Phi}_{k}(t_k)\boldsymbol{\Phi}^\dag_{k}(t_k)\boldsymbol{G}^\dag_{(k)}(t',t_k).
\nonumber \\
\eea
Let us now perform the change of integration variable $k\to s$ defined by $k=k_\epsilon(s)$. Then $\dd k^3=\dd s \left(\frac{\dd k^3_\epsilon}{\dd s}\right)$ and $t_k$ becomes $t_{k_\epsilon(s)}=k^{-1}_\epsilon[k_\epsilon(s)]=s$. One finally obtains
\bea
\boldmathsymbol{\Xi}_{t,t'}^\mathrm{QFT}\left(\vert \vec{x}'-\vec{x}\vert\right)&=&\frac{1}{6\pi^2}\ds\int^{\mathrm{min}(t,t')}_{t_\uin}\dd s
\frac{\dd k_{\epsilon}^3(s)}{\dd s} 
\mathrm{sinc}\left[k_\epsilon(s)\vert \vec{x}'-\vec{x}\vert\right]
\nonumber \\ & & \times \boldsymbol{G}_{(k_\epsilon(s))}(t,s)\boldsymbol{\Phi}_{k_\epsilon(s)}(s)\boldsymbol{\Phi}^\dag_{k_\epsilon(s)}(s)\boldsymbol{G}^\dag_{(k_\epsilon(s))}(t',s). \label{eq:covqft}
\eea
Comparing this expression with \Eq{eq:Xi:stochastic}, one clearly sees that the stochastic approach provides a good approximation of the full QFT result if the scale of coarse graining is chosen such that $\boldsymbol{G}_{(k_\epsilon(s))}(t,s)$ is sufficiently close to $\boldsymbol{G}_{(0)}(t,s)$. In other words, the stochastic calculation is valid if gradient terms can be neglected in the evolution of quantum modes at scales larger than the coarse-graining scale.

In order to make this statement more quantitative, let us compare $\boldsymbol{G}_{(0)}(t,s)$ and $\boldsymbol{G}_{(k)}(t,s)$ in more details. As explained above, they satisfy the differential equations
\bea
\label{eq:ode:G0}
\frac{\partial}{\partial t} \boldsymbol{G}_{(0)}(t,s) &=& \boldsymbol{A}_{(0)} (t)\boldsymbol{G}_{(0)}(t,s) +\boldsymbol{I} \delta(t-s)\\
\frac{\partial}{\partial t} \boldsymbol{G}_{(k)}(t,s) &=& \left[\boldsymbol{A}_{(0)} (t)+ \boldsymbol{V}_{(k)} (t)\right]\boldsymbol{G}_{(k)}(t,s) +\boldsymbol{I} \delta(t-s),
\eea
see \Eq{eq:Green:def}. By taking the difference between these two equations, one obtains
\bea
\frac{\partial}{\partial t}  \left[ \boldsymbol{G}_{(k)}(t,s) - \boldsymbol{G}_{(0)}(t,s)\right] =  \boldsymbol{A}_{(0)} (t)  \left[ \boldsymbol{G}_{(k)}(t,s) - \boldsymbol{G}_{(0)}(t,s)\right] + \boldsymbol{V}_{(k)} (t) \boldsymbol{G}_{(k)}(t,s) \, .
\eea 
Let us see this relation as a linear, first-order differential equation for $ \boldsymbol{G}_{(k)}(t,s) - \boldsymbol{G}_{(0)}(t,s)$, with a source term given by $\boldsymbol{V}_{(k)} (t) \boldsymbol{G}_{(k)}(t,s)$. By variation of constants, this equation can be formally solved as
\bea
\boldsymbol{G}_{(k)}(t,s) - \boldsymbol{G}_{(0)}(t,s) = \int_s^{t}\dd s' 
\underbrace{
\mathcal{T}\left[
\ee^{\int_{s'}^{t}\boldsymbol{A}_{(0)} (s'')\dd s''}\right]\Theta\left(t-s'\right)}_{\boldsymbol{G}_{(0)}(t,s')}
\boldsymbol{V}_{(k)} (s') \boldsymbol{G}_{(k)}(s',s),
\eea
where $\mathcal{T}$ denotes the time-ordering operator, and where one recognises the formal solution to \Eq{eq:ode:G0}, \ie $\boldsymbol{G}_{(0)}(t,s')$. This gives rise to
\bea
\label{eq:Dyson}
	\boldsymbol{G}_{(k)}(t,s)=\boldsymbol{G}_{(0)}(t,s)+\ds\int^t_s\dd s'\boldsymbol{G}_{(0)}(t,s')\boldsymbol{V}_{(k)}(s')\boldsymbol{G}_{(k)}(s',s).
\eea
This is only an implicit solution since $\boldsymbol{G}_{(k)}(t,s)$ appears on both sides of this relation. However, since both $\boldsymbol{V}_{(k)}$ and $\boldsymbol{G}_{(k)}-\boldsymbol{G}_{(0)}$ are suppressed by powers of $k^2$, this allows one to make a perturbative expansion in $k^2$ (this is the so-called Dyson series procedure) to relate $\boldsymbol{G}_{(k)}$ to $\boldsymbol{G}_{(0)}$.  The first order of this expansion is obtained by replacing $\boldsymbol{G}_{(k)}(s',s)$ by $\boldsymbol{G}_{(0)}(s',s)$ in the right-hand side of \Eq{eq:Dyson}, leading to
\bea
\label{eq:Dyson:1stOrder}
	\boldsymbol{G}_{(k)}(t,s) = \boldsymbol{G}_{(0)}(t,s)+\ds\int^t_s\dd s'\boldsymbol{G}_{(0)}(t,s')\boldsymbol{V}_{(k)}(s')\boldsymbol{G}_{(0)}(s',s)+\mathcal{O}(k^4).
\eea
The integral on the right-hand side of this expression can be computed making use of \Eq{eq:Green:massless}. Denoting $\mathcal{I}(t_1,t_2)\equiv \int_{t_1}^{t_2}\dd t/a^3(t)$, one has
\bea
\boldsymbol{G}_{(0)}(t,s')\boldsymbol{V}_{(k)}(s')\boldsymbol{G}_{(0)}(s',s) = -k^2 a(s')
\left(
\begin{array}{ccc}
\mathcal{I}(s',t) & & \mathcal{I}(s',t)\mathcal{I}(s,s')\\
1 & & \mathcal{I}(s,s')
\end{array}
\right).
\eea
If the equation-of-state parameter of the universe, $w$, is constant, then $a(t)=a_\uin (t/t_\uin)^{\frac{2}{3(1+w)}}$, and the integral $\mathcal{I}$ as well as the integration over $s'$ in \Eq{eq:Dyson:1stOrder} can be performed. This gives rise to
\bea
\label{eq:Dyson:def:M}
\boldsymbol{G}_{(k)}(t,s) - \boldsymbol{G}_{(0)}(t,s)
= \frac{k^2}{a^2(s) H^2(s)} \boldsymbol{M}(t,s)+\order{k^4},
\eea
where the matrix $\boldsymbol{M}(t,s)$ has entries
\bea
\label{eq:M00}
M_{00}\left(t,s\right)&=&\frac{ -2 }{3\left(3w+1\right)\left(w-1\right)\left(3w+5\right)}
\left\lbrace \left(3w+5\right)+3\left(w-1\right) \left[\frac{a\left(t\right)}{a\left(s\right)}\right]^{3w+1}
\right. \nonumber\\ & & \quad\quad\quad \left.
-2\left(3w+1\right)\left[\frac{a\left(t\right)}{a\left(s\right)}\right]^{\frac{3}{2}\left(w-1\right)}\right\rbrace\\
M_{10}\left(t,s\right)&=&\frac{2 a^3(s)H(s)}{3w+5}\left\lbrace 1-\left[\frac{a\left(t\right)}{a\left(s\right)}\right]^{\frac{3w+5}{2}} \right\rbrace\\
M_{01}\left(t,s\right)&=&\frac{4 a^{-3}\left(s\right) H^{-1}\left(s\right)}{3\left(3w+1\right)\left(w-1\right)\left(3w+5\right)\left(9w-1\right)}
\left\lbrace \left(3w+5\right)-\left(3w+5\right) \left[\frac{a\left(t\right)}{a\left(s\right)}\right]^{\frac{9w-1}{2}}
\right. \nonumber\\ & & \quad\quad\quad \left.
-\left(9w-1\right)\left[\frac{a\left(t\right)}{a\left(s\right)}\right]^{\frac{3}{2}(w-1)}+\left(9w-1\right)\left[\frac{a\left(t\right)}{a\left(s\right)}\right]^{3w+1}\right\rbrace\\
\label{eq:M11}
M_{11}\left(t,s\right)&=&\frac{-2 }{3\left(3w+1\right)\left(w-1\right)\left(3w+5\right)}
\left\lbrace 3\left(w-1\right)+\left(3w+5\right) \left[\frac{a\left(t\right)}{a\left(s\right)}\right]^{3w+1}
\right. \nonumber\\ & & \quad\quad\quad \left.
-2\left(3w+1\right)\left[\frac{a\left(t\right)}{a\left(s\right)}\right]^{\frac{3w+5}{2}}\right\rbrace .
\eea

With $k_\epsilon(s) = \epsilon a(s) H(s)$, \Eq{eq:Dyson:def:M} gives rise to $\boldsymbol{G}_{(k_\epsilon(s))}(t,s) - \boldsymbol{G}_{(0)}(t,s) = \epsilon^2 \boldsymbol{M}(t,s)+\order{k^4}$, so taking the difference between \Eqs{eq:Xi:stochastic} and~\eqref{eq:covqft} leads to
\bea
& & \kern-2em \boldmathsymbol{\Xi}_{t,t'}^\mathrm{QFT}\left(\vert \vec{x}'-\vec{x}\vert\right) - \boldmathsymbol{\Xi}_{t,t'}\left(\vert \vec{x}'-\vec{x}\vert\right) =
\nonumber \\ & &
\epsilon^2\int^{\mathrm{min}(t,t')}_{t_\uin}\dd s
\left\lbrace \boldsymbol{M}(t,s)\boldsymbol{\Sigma}_{s}\boldsymbol{G}^\dag_{(0)}(t',s)
+
\boldsymbol{G}_{(0)}(t,s)\boldsymbol{\Sigma}_{s}\boldsymbol{M}^\dag(t',s)
\right\rbrace + \order{\epsilon^4}\, .
\eea
In this expression, $\boldsymbol{G}_{(0)}$ is given by \Eq{eq:Green:massless}, $\boldsymbol{\Sigma}$ by \Eq{eq:Sigma:MasslessField} and $\boldmathsymbol{M}$ by \Eqs{eq:M00}-\eqref{eq:M11}. For explicitness, and in order to avoid displaying cumbersome formulas, let us evaluate this formula in the coincident configuration where $\vec{x}'=\vec{x}$ and $t=t'$. One obtains
\bea
\label{eq:Xi:QFT:Sto:Comp}
\boldmathsymbol{\Xi}_{t,t}^\mathrm{QFT}\left(0\right) - \boldmathsymbol{\Xi}_{t,t}\left(0\right) =
\epsilon^2 \frac{\left(1+3w\right)^{2\nu-1} \epsilon^{3-2\nu}\Gamma^2(\nu)}{4\pi^3 w }
\boldmathsymbol{N}(t) + \order{\epsilon^4}\, ,
\eea
where the matrix $\boldsymbol{N}(t)$ has entries
\bea
\label{eq:N00}
N_{00}&=&\frac{H^2}{9 \left(1+9w\right)\left(1-9w\right)\left(1-w\right)}\left[\left(1-w\right)\left(9w-1\right)+\left(1-9w\right)\left(1+9w\right)\left(\frac{a}{a_\uin}\right)^{6w} 
\right. \nonumber \\ & & \left. 
+18w\left(w-1\right)\left(\frac{a}{a_\uin}\right)^{9w+1}+8w\left(1+9w\right)\left(\frac{a}{a_\uin}\right)^{\frac{3}{2}\left(3w+1\right)}\right]\\
N_{01}&=&N_{10}=\frac{a^3 H^3}{6\left(9w-1\right)\left(9w+1\right)}\left[\left(w+1\right)\left(9w-1\right)+\left(1-9w\right)\left(1+9w\right)\left(\frac{a}{a_\uin}\right)^{6w}
\right. \nonumber \\ & & \left. 
+12 w \left(3w-1\right) \left(\frac{a}{a_\uin}\right)^{9w+1}
+4w\left(9w+1\right)\left(\frac{a}{a_\uin}\right)^{\frac{3}{2}\left(3w+1\right)}
\right]\\
N_{11}&=& \frac{w-1}{4\left(1+9w\right)} a^6H^4\left[\left(1+3w\right)-\left(1+9w\right)\left(\frac{a}{a_\uin}\right)^{6w}+6w \left(\frac{a}{a_\uin}\right)^{9w+1}\right]\, .
\label{eq:N11}
\eea
This needs to be compared with the stochastic result we have been using, namely
\bea
\kern-2em
\boldmathsymbol{\Xi}_{t,t}\left(0\right) &=& 
\int_{t_\uin}^t \dd s
\boldsymbol{G}_{(0)}(t,s)\boldsymbol{\Sigma}_{s}\boldsymbol{G}^\dag_{(0)}(t,s) \nonumber \\
& = &
\frac{\left(1+3w\right)^{2\nu}\epsilon^{3-2\nu}\Gamma^2(\nu)}{16\pi^3 w}
\left[1-\left(\frac{a}{a_\uin}\right)^{6w}\right]
\left(
\begin{array}{cc}
\frac{H^2}{3} & \frac{w-1}{2}a^3 H^3 \\
 \frac{w-1}{2}a^3 H^3 & \frac{3}{4}\left(w-1\right)^2 a^6 H^4
\end{array}
\right)\, .
\label{eq:Xi:sto:explicit}
\eea
Let us note that in the square bracketed terms of \Eqs{eq:N00}-\eqref{eq:N11}, soon after the onset of the contracting phase, since $a/a_\uin \ll 1$, the dominant term is the one with the smallest power of $a/a_\uin$. Assuming that $w\in ]-1/3,1[$, this power is either $0$, when $w>0$, or $6w$ when $w<0$. Indeed, $3w+1>0$ for $w>-1/3$ so $3/2(3w+1)$ is never the smallest power index, and $9w+1>6w$ for $w>-1/3$ so $9w+1$ is never the smallest power index either. This means that, as in \Eq{eq:Xi:sto:explicit}, the result either scales as $(a/a_\uin)^0$ or as $(a/a_\uin)^{6w}$. It is then easy to see that the prefactors appearing in \Eqs{eq:Xi:QFT:Sto:Comp} and~\eqref{eq:Xi:sto:explicit} are the same, up to overall constants of order one, and up to the (crucial) additional $\epsilon^2$ factor appearing in \Eq{eq:Xi:QFT:Sto:Comp}. One concludes that, provided $\epsilon\ll 1$, the stochastic result provides a good approximation to the full QFT result. This has been derived here for the coincident configuration of the field correlators, but this can be easily extended to the non-coincident configuration, and to the two-point function of the anisotropic stress since the two are simply related via \Eq{eq:Piji:Piij:Xi''}. This validates the use of the stochastic formalism to perform the calculation presented in this work.

\bibliographystyle{JHEP}
\bibliography{StochaAnisotropy}
\end{document}